\documentclass[12pt]{article}

\usepackage[margin=1in]{geometry}
\usepackage{setspace}
\IfFileExists{lmodern.sty}{\usepackage[T1]{fontenc}\usepackage{lmodern}}{}
\usepackage[expansion=false]{microtype}
\usepackage{parskip}

\usepackage{amsmath,amssymb,amsthm,mathtools,bm}
\theoremstyle{plain}

\newtheorem{proposition}{Proposition}
\newtheorem{lemma}{Lemma}

\theoremstyle{definition}
\newtheorem{assumption}{Assumption}

\theoremstyle{remark}

\usepackage{booktabs,multirow,array,dcolumn}
\usepackage{graphicx}
\usepackage{caption}
\usepackage[authoryear,round]{natbib}
\usepackage{xcolor}
\usepackage[colorlinks=true,linkcolor=blue!60!black,citecolor=blue!60!black,urlcolor=blue!60!black]{hyperref}

\newif\ifdraft\draftfalse

\newcommand{\E}{\mathbb{E}}
\newcommand{\Var}{\operatorname{Var}}

\newcommand{\ATT}{\mathrm{ATT}}
\newcommand{\ATU}{\mathrm{ATU}}
\newcommand{\ATE}{\mathrm{ATE}}
\newcommand{\FDR}{\mathrm{FDR}}
\newcommand{\FOR}{\mathrm{FOR}}

\newcommand{\Dh}{\widehat{D}}

\newcommand{\FDRbar}{\overline{\FDR}}

\title{\Large\textbf{Who Is the Regression About?\\
Auditing AI-Labeled Treatments When Effects Differ}}
\author{Ayman Farahat\thanks{Yahoo. Email: \texttt{ayman.farahat@yahooinc.com}. No Yahoo data were used in this paper, and the views expressed are the author's and not necessarily those of Yahoo. An AI assistant (Claude, Anthropic) helped write the code, carried out the first-pass reading of audit records under the protocol of Section~\ref{sec:procedure}, and helped draft the text; the author checked the results and made or reconciled every audit call.}}
\date{November 2026}

\newcounter{boxc}
\begin{document}
\maketitle

\begin{abstract}
\noindent Researchers increasingly regress outcomes on treatments labeled by large language models. When effects differ, OLS averages the effect on the treated (ATT) and on the untreated (ATU), weighting the smaller group more. A label's bias then has two parts: classical attenuation from misclassification, and a selective term when its errors are correlated with the outcome. In LinkedIn postings, the model misses better-paid remote jobs: a precision check passes, yet the remote pay gap is understated by nearly 40 percent, a bias that more data cannot remove. A stratified, blind audit of both labeled groups corrects it and recovers both effects.
\end{abstract}


\thispagestyle{empty}
\newpage
\setcounter{page}{1}

\section{Introduction}
\label{sec:intro}

Large language models have made it cheap to turn text into variables. A researcher who once hand-coded a few hundred documents can now label a million: which job postings offer remote work, which firms mention a new technology, which court opinions cite a precedent. The label then enters a regression as the treatment, and the regression is estimated on the full labeled sample, with standard errors to match. This paper is about what that regression estimates, why it is biased in a way that more data cannot fix, and what to do about it.

Start with what the regression is trying to recover. A binary treatment has an effect that differs across units, and there are two natural averages of it. The average effect on the treated (ATT) is the effect for the units that actually received the treatment: for job postings, what going remote did to the pay of the jobs that are remote. The average effect on the untreated (ATU) is the effect the treatment would have on the units that did not receive it: what going remote would do to the pay of the jobs that are on site. They differ whenever the treatment is more common where its effect is larger or smaller; in the LinkedIn data the remote pay premium is largest in industries where remote jobs are rare, so the ATU (0.21 log points) exceeds the ATT (0.14). A regression of the outcome on the treatment and controls reports neither. \citet{Sloczynski2022OLS} shows that the OLS coefficient, call it $\beta$, is a weighted average of the two, $\beta=w_1\,\ATT+w_0\,\ATU$, and that the weights run against the group sizes: the smaller the treated group, the larger the weight on its effect. When 13 percent of postings are remote, 81 percent of the weight falls on the effect for remote postings. The regression is about the minority.

Now let the treatment be a label. A classifier or a language model makes two kinds of error. False positives put untreated units into the flagged group; misses leave treated units among the controls. \citet{Aigner1973Binary} showed that either error attenuates the coefficient: the flagged and unflagged groups are each mixtures of treated and untreated units, and the gap between two mixtures is smaller than the gap between the pure groups. What matters for the regression is not the classifier's error rate but the share of each group that is contaminated, and that share is largest exactly where the treatment is rare, because a small flagged group is fed by false positives from a very large untreated pool, and the misses of a small treated group are its most consequential members. So the label hurts the minority most, and the minority is where \citeauthor{Sloczynski2022OLS} says the regression's weight lies. The two results compound in a second way as well. By missing treated units the label makes the minority look even smaller, and a smaller minority gets a larger weight: the label attenuates the effect on the treated and then tells OLS to lean on it harder. I show that the bias of the OLS coefficient on a label decomposes exactly into these two channels, an effects channel, in which each group's effect is replaced by its attenuated version, and a weights channel, in which the weights shift toward the damaged effect, and that the attenuation itself carries an extra term when the label's errors are \emph{selective}, that is, when the units it gets wrong differ in the outcome from the units it gets right, as a language model's errors can be when it reads the same text that describes the job.

Neither channel depends on the sample size. The errors are a property of the label, not of the draw, so labeling ten times as many records leaves the bias where it is while the standard error shrinks by two thirds. \citet{Battaglia2024Inference} draw the consequence: the confidence interval of the naive regression tightens around a number that is wrong, and its coverage of the truth falls to zero as the sample grows. At the scale at which language models label data this is the regime that matters. In the LinkedIn data the naive 95 percent interval covers the true $\beta$ in none of 200 resampled samples at the size of the sample itself.

The paper's contribution is a procedure that a researcher can follow to recover from this: \emph{audit the label, correct the cells, choose the estimand}. Cells are bins of the label's propensity score, the probability that the label flags a unit given the controls. The audit reads a small sample of records drawn from every cell and from both the flagged and the unflagged columns, blind to the outcome, with two readers and a written rule. The correction replaces each cell's labeled moments by the full-sample label average plus the audit's estimate of the gap to the truth, a prediction-powered difference estimator \citep{Angelopoulos2023PPI} applied cell by cell; because it uses audited differences rather than estimated error rates, it remains valid when the label's errors depend on the outcome. The researcher then reports the ATT, the ATU and the regression coefficient $\beta$, together with \citeauthor{Sloczynski2022OLS}'s weights, and says which one answers the question. I give the estimator's properties, a budget rule for the audit size, and an open-source implementation, including the blind audit sheets, a labeling tool and a protocol for using a language model as one of the two readers.

I apply the procedure to the remote-work pay gap in 35,618 LinkedIn job postings labeled by a 7-billion-parameter language model, and the application yields four lessons that carry beyond it. First, the naive OLS coefficient $\beta$ on the label is 0.086 log points against a corrected 0.14, nearly 40 percent too low, and the ATT and the ATU are understated by about the same proportion. Second, the bias sits in the label's misses, not its false positives. The label's false-positive rate is below one percent in every cell, and 86 percent of the postings it seems to flag wrongly are remote by their text (the benchmark field is what is wrong), so a precision check on flagged records, the usual validation, would have found a highly precise label. Yet the label misses 55 percent of remote postings, and the misses are selective: within the same cell, the remote postings it misses pay 8 log points more than those it catches, because it catches remote jobs that say so at the top of a short posting and misses those that say so once in a long one, and long postings are for better-paid jobs. Keeping the label's number of misses in every cell but choosing them at random recovers two thirds of the bias, so most of the damage comes from \emph{which} units are missed, not how many. Removing the misses from the label removes the bias entirely; removing the false positives leaves it intact. The classical formula would not have predicted this, because the misses are a small share of the column they land in; they bite because they are selective and because they shrink the apparent minority. Because this rests on one dataset, I also show when it holds (Proposition~\ref{prop:regime}, Figure~\ref{fig:regime}): with random errors and a rare treatment, false positives carry the bias, as the classical analysis says; once the missed units differ from the caught ones in the outcome by about half the effect, misses take over, even with one or two misses per false positive. A pilot that reads both columns tells the researcher which case applies. Measuring false positives is cheap, because the flagged column is small; measuring misses is expensive, because they hide in the column that holds nearly everything, and that asymmetry is what the audit design has to overcome. Third, the regression is about the minority, as \citeauthor{Sloczynski2022OLS} predicts: after correction the weight on the treated is 0.84, and the pay gap for remote postings (0.12 to 0.13) is about 40 percent smaller than the gap that would apply to postings that are not remote (0.20 to 0.21). Fourth, the audit itself must be designed with care: its uncertainty is dominated by the largest unflagged stratum, where one reading can stand for 300 records; in the LinkedIn data a person reading with the salary in view coded remote status in a way that tracked pay about three times as steeply as when reading blind, a suggestive but not conclusive difference; and masking whole sentences that mention pay deletes the treatment cue in 6 to 12 percent of postings.

\paragraph{When the results apply.} The setting is common and simple. Economists now routinely impute a binary variable with a classifier and use it as a regressor: the gender of borrowers and the ethnicity of donors from their names \citep{GoldsmithPinkhamShue2023Gender,Bursztyn2024Immigrant}, the remote status of jobs from their postings \citep{Hansen2023Remote}, the gender and tone of seminar speakers from recordings \citep{Dupas2026Seminars}; \citet{Christensen2026Bootstrap} collect further examples. A researcher regresses an outcome on a binary treatment and a set of controls, by OLS, and the treatment is not observed but produced by a classifier: a language model reading text, a supervised model, a keyword rule. Three conditions make the problem bite, and each is typical of LLM labels. The first is that the effect differs across units and the treatment is unbalanced across the controls, so the ATT and the ATU differ and OLS weights them unequally; if the effect were the same for everyone there would be no weights to distort, though the attenuation and its bias would remain. The second is that the label errs, and in particular that it misses treated units; a label with few misses and a modest false discovery rate does far less harm. The third is that the sample is large, as LLM-labeled samples usually are, so that the bias dominates the standard error and the confidence interval misleads. Language models also make a fourth feature likely, which the classical analysis rules out: errors that are related to the outcome, because the model reads the same text that describes the job. The results do not require any particular classifier, and the procedure applies equally to labels from any source; what it requires is that a sample of records can be read and the truth recorded, and that the outcome can be hidden from the people and models that read them.

\paragraph{The intuition for the audit.} The usual check of a label is its precision: read a sample of the records it flags and count how many are right. Precision measures false positives, and false positives are the easy error to find, because the flagged column is small. This paper shows that this check looks in the wrong place. The regression's slope is mostly the effect on the treated (the minority gets the larger weight), and that effect is estimated as the gap between the treated and their counterfactual, the untreated units that look like them. Misses damage both sides of that gap. They take treated units out of the flagged column, and because the units a language model misses are not random, here the better-paid remote jobs, the flagged column's mean falls; and they put those same units into the unflagged column, among the counterfactuals, where they raise its mean exactly where the treated sit. They also make the treated group look smaller, which raises its weight in the slope. So the audit must measure misses, not only precision, and misses are hard to find: they are a small share of the large unflagged column. That is why the audit is stratified, reading the unflagged column in every cell and hardest where the treated are most likely to hide, and why it uses the audited differences directly rather than estimated error rates.

\paragraph{Related literature.} I build on what OLS with controls estimates under heterogeneity \citep{Angrist1998Estimating,Sloczynski2022OLS}, on related work on weighted and contaminated estimands \citep{GoldsmithPinkhamHullKolesar2024Contamination,MogstadTorgovitsky2024IV,PoirierSloczynski2024Internal}, and on the econometrics of a misclassified binary regressor \citep{Aigner1973Binary,Bollinger1996Bounding,Mahajan2006Misclassification,Lewbel2007Misclassification,Hu2008Misclassification}, which mostly identifies error rates from instruments or repeated measures under a common effect; with heterogeneity, \citet{BattistinSianesi2011Misclassified} bound the effect under selection on observables. Closest in spirit, \citet{DentehKedagni2022DiD} and \citet{NegiNegi2025DiD} show that a misclassified treatment changes the difference-in-differences estimand; \citet{NegiNegi2025DiD} correct it with a two-step model of misreporting in the spirit of \citet{Heckman1979Sample}, identified by an exclusion restriction and joint normality. I correct with data instead, from an audit, which requires no model of the label's errors and remains valid when they depend on the outcome. The literature on inference with AI- or ML-generated variables \citep{FongTyler2021Predictions,Angelopoulos2023PPI,Egami2023DSL,Battaglia2024Inference,Kluger2025Imputed,Carlson2026Unifying,Christensen2026Bootstrap,ChristensenHansen2026Guide} uses a validated subsample to correct a regression with a predicted variable, and the auxiliary-data and two-phase sampling literature supplies the estimator and the allocation \citep{Neyman1934Two,ChenHongTamer2005Auxiliary}. \citet[Proposition~3 and Appendix~C.2]{LudwigMullainathanRambachan2026LLM} give the validation-sample correction for a regression with an LLM label on either side, with a simple random validation sample and no treatment-effect structure, and they note that the bias depends on how the label's errors covary with the economic variables without observing those errors; my contribution is to the structure, not the estimator, and I document the mechanism they posit: in the LinkedIn data the errors are concentrated among the better-paid treated units, and that selectivity, not the error count, carries most of the bias. Two recent papers raise related concerns about LLM annotations without measuring the misses: \citet{Overgoor2025Wrong} show, in television advertising, that a noisy human benchmark can pull corrections toward the wrong target, and \citet{Camuffo2026Variance} note that annotation errors correlated with the study variables can bias regressions even when accuracy is high. I apply that logic to treatment-effect targets under heterogeneity: correcting the coefficient once recovers the coefficient one would get with the true treatment, which is still a weighted mix; correcting the cells lets the researcher choose the weights, and the cell structure tells the researcher where the audit must look.

\paragraph{Relation to the closest work.} Two papers study the same regression and, in part, the same data. \citet{Battaglia2024Inference} derive the bias of OLS on a machine-generated binary regressor and an analytic correction; \citet{Christensen2026Bootstrap} show how to bootstrap it, with a coupled-label bootstrap that draws true and imputed labels jointly, and apply it to the San Diego remote-work postings of \citet{Battaglia2024Inference}. Both need only the label's false-positive and false-negative rates, from an external sample that records the truth and the label but not the outcome. That makes them the right tools when the truth cannot be linked to the outcome, as when a name-based classifier is validated on a voter roll. Three of their maintained conditions are ones the setting here relaxes, and the LinkedIn data show why that matters. First, the coefficient on the label is common to all units, or varies only through known interactions with observed covariates, so the target is a single coefficient. Under heterogeneity that coefficient is \citeauthor{Sloczynski2022OLS}'s mix of the ATT and the ATU, and a correction recovers the mix rather than either effect; interacting the label with cell dummies would bring their framework close to the cells used here, but still under the next two conditions. Second, the label's errors must be uncorrelated with the regression error up to terms smaller than the sampling error \citep[Assumption~3(iii)]{Christensen2026Bootstrap}, which rules out the selective errors of Lemma~\ref{lem:gap}. In the LinkedIn data the remote postings the language model misses pay 8 log points more than those it catches in the same cell, and that selectivity carries about two-thirds of the bias. Third, a calibration condition \citep[Assumption~3(v)]{Christensen2026Bootstrap} implies, in the additive model, that the label makes as many false positives as misses. In the application they accordingly set the false-negative rate equal to the false-positive rate of 0.009 that \citet{Battaglia2024Inference} measure by reading the flagged postings, and double it as a robustness check. The condition is natural for a classifier trained on the population it labels, and the DistilBERT classifier they use is consistent with it on the LinkedIn postings: against the employer's field it makes about 2,700 false positives and 2,500 misses. A zero-shot language model need not be calibrated in this sense, and the one used here is not: against the same field it makes about 300 false positives and 2,600 misses. Because the audit proposed here reads records from both columns together with their outcomes, it measures the misses and their selectivity instead of assuming them. The price is that the truth must be readable from the same records as the outcome, which is the case for labels read from text. \citet{ChristensenHansen2026Guide} give practitioners a guide and software for these corrections, and note that complete validation data are often unavailable because the latent variable is never observed, as for sentiment or uncertainty; there, a correction from error rates is the only option. Where the truth can be read, as for remote work, a small audit of both columns tells the researcher whether a correction from the false-positive rate alone is enough (Section~\ref{sec:measure}).

\paragraph{What is borrowed and what is new.} To be plain about it: the correction I apply is the validation-sample correction of prediction-powered inference and two-phase sampling, as \citet{LudwigMullainathanRambachan2026LLM}, \citet{Fisch2024StratPPI} and \citet{Kluger2025Imputed} present it, applied within cells, and more efficient estimators of causal effects with a mismeasured treatment and a validation sample exist \citep{Barnatchez2025TwoPhase}; the decomposition of the OLS coefficient into group effects and weights is \citeauthor{Sloczynski2022OLS}'s; the attenuation of a misclassified binary regressor is \citeauthor{Aigner1973Binary}'s; and the observation that the bias survives a growing sample is \citeauthor{Battaglia2024Inference}'s. What is new here is the structure that a binary treatment under heterogeneous effects imposes on all four. The regression leans on the minority; a label's two errors land in two columns and the base rates make the minority's columns the most contaminated; the misses matter through their selectivity, weighted by the miss rate, and through the weights they shift, so that removing them removes the bias while removing the false positives does nothing; the coefficient the standard correction recovers is therefore the wrong target, and the cells have to be corrected instead so that the researcher can choose the estimand; and the audit has to be stratified by cell and column and read blind, because a simple random validation sample contains almost no misses and a reader who sees the outcome codes differently. The recall-not-precision lesson, the budget rule and the reading protocol follow from that structure.

\paragraph{Roadmap.} Section~\ref{sec:bias} shows why the bias does not wash out. Section~\ref{sec:procedure} gives the procedure and its properties. Section~\ref{sec:application} applies it to LinkedIn postings, and Section~\ref{sec:size} asks how large the audit must be. Section~\ref{sec:conclusion} concludes. Proofs are in the appendix.

\section{Why the bias does not wash out}
\label{sec:bias}

This section builds the argument in three steps, each resting on a known result, and ends with what the argument implies for measuring a label. \citet{Sloczynski2022OLS} shows that the OLS coefficient is a weighted average of two treatment effects, with the larger weight on the smaller group (Section~\ref{sec:ols}). \citet{Aigner1973Binary} shows that a misclassified binary regressor attenuates the coefficient; I show that the attenuation is strongest in the small group, and that a language model's errors add a term the classical analysis rules out, because the units it gets wrong differ in the outcome from the units it gets right (Section~\ref{sec:label}). The two results compound: the label damages the minority's effect and then shifts weight onto it (Section~\ref{sec:channels}). The practical consequence is that the label's precision, the usual check, is the wrong quantity; what matters is its recall in each cell and which units it misses, and I show that the misses in the LinkedIn data fall on the better-paid treated units and carry the whole of the bias (Section~\ref{sec:measure}). Finally, \citet{Battaglia2024Inference} show that none of this shrinks as the sample grows while the standard error does (Section~\ref{sec:trap}).

\subsection{Two treatment effects}

Unit $i$ has an outcome $Y_i$, controls $X_i$, a binary treatment $D_i$ and two potential outcomes, $Y_i(1)$ if treated and $Y_i(0)$ if not. Only one of them is observed. In the application a unit is a job posting, the outcome is log salary, and the treatment is that the job is remote. The treatment effect for posting $i$ is $Y_i(1)-Y_i(0)$: what the job would pay remote minus what it would pay on site. It differs across jobs, and the question is whose effects a regression averages.

Partition the controls into \emph{cells} $w$ (industry groups, say, or bins of a propensity score, as below), with shares $s(w)=\Pr(W=w)$, treatment rates $p(w)=\Pr(D=1\mid w)$, mean potential outcomes $\mu_d(w)=\E[Y(d)\mid w]$ and cell effects $\tau(w)=\mu_1(w)-\mu_0(w)$. I assume:

\begin{assumption}[Ignorability within cells]\label{as:ign2}
$D\perp\big(Y(1),Y(0)\big)\mid W$.
\end{assumption}

Under Assumption~\ref{as:ign2}, the mean outcome of treated units in a cell identifies $\mu_1(w)$, that of untreated units identifies $\mu_0(w)$, and their difference identifies $\tau(w)$. The three standard averages of these cell effects differ only in whom they average over:
\begin{equation}\label{eq:estimands}
\ATT=\frac{\sum_w s(w)\,p(w)\,\tau(w)}{\sum_w s(w)\,p(w)},\qquad
\ATU=\frac{\sum_w s(w)\{1-p(w)\}\,\tau(w)}{\sum_w s(w)\{1-p(w)\}},\qquad
\ATE=\sum_w s(w)\,\tau(w).
\end{equation}
The \emph{average treatment effect on the treated} (ATT) weights each cell by its number of treated units: it is the average effect for the jobs that are in fact remote, and answers ``what did going remote do to the pay of the jobs that went remote?'' The \emph{average treatment effect on the untreated} (ATU) weights each cell by its number of untreated units: it is the average effect for the jobs that are not remote, and answers ``what would happen to pay if these jobs went remote?'' The ATE weights by cell size.

The ATT and ATU differ when two things hold at once: the effect $\tau(w)$ varies across cells, and the treatment rate $p(w)$ varies across cells, so that the treated and the untreated sit in different cells. Both hold here. Remote jobs cluster in cells where remote is common (software, finance, professional services); on-site jobs cluster where it is rare (hospitals, retail, manufacturing). In the cells where remote is rare, the few remote jobs pay much more than the on-site ones, so $\tau(w)$, the gap in mean log pay between remote and on-site postings within the cell, is large where $p(w)$ is small: 0.42 log points in the lowest cell against 0.05 in the highest, declining across the cells though not strictly (Table~\ref{tab:errors}, last columns). Two things lie behind the decline. One is composition that the cells do not remove: each cell pools several industries and experience levels, and in the low cells the remote postings are disproportionately senior (10 percent against 4 percent of on-site postings in the lowest cell). Comparing remote and on-site postings within the same industry, experience level and work type inside each cell lowers the gap to 0.24 in the lowest cell and 0.01 in the highest, so composition accounts for part of the decline but not most of it. The other is the nature of the jobs: where remote work is rare (retail, government, manufacturing), the remote postings are for different jobs than the on-site ones, corporate and back-office roles rather than store, plant or counter work, while where remote work is common (software, insurance, IT services) remote and on-site postings are for much the same jobs and pay much the same. These are conditional pay gaps, and calling them effects requires Assumption~\ref{as:ign2}; what matters for this paper is that they differ across cells, which is what makes the ATT and the ATU differ. Averaging the large gaps over the many on-site jobs gives an ATU above the ATT. With the employer's remote field as the treatment, the ATT is $0.140$ log points and the ATU $0.214$.

The policy question usually concerns the ATU. Whether more jobs should go remote is a question about the jobs that are not remote now. A regression that mostly reports the ATT understates the answer by a third here, before any label error enters.

\subsection{What OLS estimates: the smaller group gets the larger weight}
\label{sec:ols}

Regress $Y$ on $D$ and cell dummies. The coefficient is the variance-weighted average of the cell effects, $\beta=\sum_w s(w)p(w)\{1-p(w)\}\tau(w)\big/\sum_w s(w)p(w)\{1-p(w)\}$ \citep{Angrist1998Estimating}: cells where the treatment is split evenly count most, and cells where almost everyone or almost no one is treated count little. \citet{Sloczynski2022OLS} rewrites the same coefficient as a convex combination of the two group effects,
\begin{equation}\label{eq:sloczynski}
\beta \;=\; w_1\,\tau_1 + w_0\,\tau_0,\qquad w_1=\frac{(1-\rho)\,V_0}{\rho\,V_1+(1-\rho)\,V_0},\quad w_0=1-w_1,
\end{equation}
where $\rho=\Pr(D=1)$ is the treated share and $V_d$ is the variance of the propensity score $p(w)$ among units with $D=d$. The two effects $\tau_1$ and $\tau_0$ come from a three-step construction \citep[eqs.~5--8]{Sloczynski2022OLS}: compute the propensity score $p(w)$; fit two lines, the linear projection of $Y$ on $p(w)$ among treated units and among untreated units; and read the vertical gap between the lines, averaged over the propensity scores of the treated for $\tau_1$ and over those of the untreated for $\tau_0$. When the two lines are the conditional means, $\tau_1=\ATT$ and $\tau_0=\ATU$; when the conditional means are not linear in $p(w)$, the lines approximate them and $\tau_1$, $\tau_0$ differ from the cell averages~\eqref{eq:estimands} by what \citeauthor{Sloczynski2022OLS} calls the bias from nonlinearity. I report both throughout.

The weights in~\eqref{eq:sloczynski} are the point. The weight on the treated group's effect, $w_1$, is decreasing in the treated share $\rho$: the smaller the group, the larger the weight on its effect. When 13 percent of postings are remote, OLS puts a weight of $w_1=0.81$ on the effect for remote postings and $0.19$ on the effect for the 87 percent that are not. Why is the regression about the minority? The counterfactual for each group is read off the other group's line: the treated units' $\tau_1$ is the gap between their own line and the untreated line, evaluated where the treated sit. A small treated group sitting at high propensity scores is compared with the untreated units at the same scores, and OLS trusts that comparison more than the reverse one, because the untreated line is estimated from many units while the treated line is estimated from few and is noisy where the untreated sit. So with the true treatment the regression reports $\beta=0.150$, close to the ATT of $0.140$ and far from the ATU of $0.214$. This is \citeauthor{Sloczynski2022OLS}'s result and not mine; what follows is what a label does to it.

\subsection{What a label does, and why it hits the minority hardest}
\label{sec:label}

Now replace $D$ by a label $\Dh$ from a classifier or a language model. Call a unit \emph{treated} if $D=1$ and \emph{flagged} if $\Dh=1$. The label makes two kinds of error, and they land in different places.

\paragraph{Two errors, two columns.} The regression compares two \emph{columns} of units, those with $\Dh=1$ and those with $\Dh=0$; with the true treatment these are the treated and the untreated, with a label they are the flagged and the unflagged. A \emph{false positive} is an untreated unit that is flagged: it moves out of the untreated column and into the flagged one. A \emph{miss} (false negative) is a treated unit that is not flagged: it stays in the unflagged column, where the regression treats it as a control. A column is \emph{contaminated} when it holds units that do not belong to it: the flagged column is contaminated by false positives, the unflagged column by misses. The regression never sees the errors; it sees the two contaminated columns and takes the gap between their means.

The classifier is described by its error \emph{rates}, the false-positive rate $a(w)=\Pr(\Dh=1\mid D=0,w)$ (the share of untreated units it flags) and the miss rate $m(w)=\Pr(\Dh=0\mid D=1,w)$ (the share of treated units it fails to flag). The columns are described by their \emph{contamination}, the share of each column that is on the wrong side:
\begin{equation}\label{eq:fdrfor}
\begin{aligned}
\FDR(w)&=\Pr(D=0\mid \Dh=1,w)=\frac{\{1-p(w)\}\,a(w)}{\hat p(w)},\\
\FOR(w)&=\Pr(D=1\mid \Dh=0,w)=\frac{p(w)\,m(w)}{1-\hat p(w)},
\end{aligned}
\end{equation}
where $\hat p(w)=p(w)\{1-m(w)\}+\{1-p(w)\}a(w)$ is the flag rate in the cell. The \emph{false discovery rate} $\FDR(w)$ is the share of the flagged column that is not treated; the \emph{false omission rate} $\FOR(w)$ is the share of the unflagged column that is treated. Rates and contaminations are linked by the base rates: the numerator of each contamination is an error rate times the size of the column the errors come from, and the denominator is the size of the column they land in.

Table~\ref{tab:twocells} makes the two pairs of numbers concrete with two cells of the LinkedIn data. Each cell is a $2\times2$ table of postings, by true status (remote or not, the columns of the table) and label (flagged or not, the rows). The errors are the same two boxes in both readings: the false positives (flagged but not remote) and the misses (remote but unflagged). The difference is what they are divided by. The \emph{error rates} divide each error by the true group it comes from, and describe the classifier: in cell 1 the model flags 10 of 3,502 non-remote postings (a false-positive rate of 0.3 percent) and misses 44 of 60 remote ones (73 percent). The \emph{contaminations} divide each error by the labeled column it lands in, and describe what the regression compares: of the 26 postings the model flags in cell 1, 10 are not remote, so the flagged column is 38 percent contaminated (FDR); of the 3,536 it leaves unflagged, 44 are remote, so the unflagged column is 1.2 percent contaminated (FOR). Read this way, FDR answers ``if I take a flagged posting, how likely is it not to be remote?'' and FOR answers ``if I take an unflagged posting, how likely is it to be remote?''. In cell 10 the picture reverses: the flagged column is only 7 percent contaminated, while the 743 misses make up 26 percent of the unflagged column.

\begin{table}[htbp]\centering\small
\caption{Error rates and column contamination: two cells as $2\times2$ tables}\label{tab:twocells}
\resizebox{\linewidth}{!}{%
\begin{tabular}{lrrr@{\hskip 24pt}lrrr}\toprule
\multicolumn{4}{c}{Cell 1 (remote is rarest)} & \multicolumn{4}{c}{Cell 10 (remote is most common)}\\
\cmidrule(r{18pt}){1-4}\cmidrule{5-8}
 & Remote & Not remote & Column & & Remote & Not remote & Column\\\midrule
Flagged & 16 & \textbf{10} & 26 & Flagged & 660 & \textbf{50} & 710\\
Unflagged & \textbf{44} & 3,492 & 3,536 & Unflagged & \textbf{743} & 2,109 & 2,852\\
Group & 60 & 3,502 & 3,562 & Group & 1,403 & 2,159 & 3,562\\\midrule
\multicolumn{4}{l}{Classifier error rates (divide by the group, a column of the table)} & & & & \\
\quad False-positive rate & \multicolumn{3}{l}{$10/3{,}502=0.3\%$} & & \multicolumn{3}{l}{$50/2{,}159=2.3\%$}\\
\quad Miss rate & \multicolumn{3}{l}{$44/60=73\%$} & & \multicolumn{3}{l}{$743/1{,}403=53\%$}\\
\multicolumn{4}{l}{Column contamination (divide by the regression's column, a row of the table)} & & & & \\
\quad FDR, flagged column & \multicolumn{3}{l}{$10/26=38\%$} & & \multicolumn{3}{l}{$50/710=7\%$}\\
\quad FOR, unflagged column & \multicolumn{3}{l}{$44/3{,}536=1.2\%$} & & \multicolumn{3}{l}{$743/2{,}852=26\%$}\\
\bottomrule\end{tabular}}
\par\smallskip\parbox{0.95\linewidth}{\footnotesize Notes: counts of postings in the lowest and highest cells of Table~\ref{tab:errors}; ``remote'' is the employer's field; ``flagged'' is the language-model label. Bold cells are the errors: false positives (flagged, not remote) and misses (remote, unflagged). The same errors are divided by different totals: the error \emph{rates} by the size of the true group the errors come from, the \emph{contaminations} by the size of the labeled column the regression compares.}
\end{table}

Equation~\eqref{eq:fdrfor} is where the minority gets hurt. When the treatment is rare, $1-p(w)$ is large and $\hat p(w)$ is small, so even a tiny false-positive rate produces a large $\FDR(w)$: the flagged column is a small pool fed by false positives from a very large one. The mirror statement holds for misses: a large miss rate produces only a small $\FOR(w)$, because the misses, however many of the treated they are, are poured into the column that holds nearly everyone. The same classifier, judged by its error rates, may be equally good everywhere; judged by what it does to the regression's columns, it is far worse where the treated are few.

\paragraph{Table~\ref{tab:errors}: the language-model label, cell by cell.} Table~\ref{tab:errors} shows all of this for the language-model label, taking the employer's remote field as the truth for now (the audit in Section~\ref{sec:application} revises that). The rows are the ten cells, deciles of the label's propensity score (Section~\ref{sec:procedure}), each with about 3,560 postings; cell 1 is where the label, and the employer field, say remote is rarest, cell 10 where it is most common. The columns are computed from the full sample. \emph{Share remote} gives the true treatment rate $p(w)$, the share of the cell's postings the employer marks remote, and the flag rate $\hat p(w)$, the share the model flags; the model flags about half as many postings as the field marks in every cell (1.7 against 0.7 percent in cell 1, 39 against 20 in cell 10). \emph{Classifier error} gives the two rates: the false-positive rate, the share of the cell's non-remote postings the model flags, is between 0.2 and 2.3 percent; the miss rate, the share of the cell's remote postings it does not flag, is between 50 and 75 percent, with no clear gradient across cells. By these two numbers the model is a cautious classifier: it almost never flags a posting the field calls non-remote, and it fails to flag half of those the field calls remote. \emph{Column contamination} translates the rates into what the regression sees. The FDR, the share of the model's flags in the cell that are not remote by the field, falls from 38 percent in cell 1 to 7 percent in cell 10: the same sub-one-percent false-positive rate, applied to a cell that is 98 percent non-remote, overwhelms a flagged column that holds 0.7 percent of the cell. The FOR, the share of the cell's unflagged postings that are remote, rises from 1 to 26 percent across the cells, and is below 14 percent in nine of the ten; the misses are half of all remote postings, but in the unflagged column they are a minority. \emph{Pay gap} gives the quantity the regression is built from, within each cell: the true gap $\tau(w)$, the mean log pay of remote minus non-remote postings, and the labeled gap $\hat\tau(w)$, flagged minus unflagged. The true gap is largest where remote is rarest, 0.42 log points in cell 1 against 0.05 in cell 10, which is the heterogeneity that makes the ATU exceed the ATT. The labeled gap is smaller than the true gap in every cell but one, by a factor that is largest in the low-propensity cells (0.14 against 0.42 in cell 1).

Three conclusions follow. First, the two error rates are flat across cells but the contaminations are not; what the regression suffers is the contamination, so a validation that reports an error rate, or a single pooled accuracy, says little about the regression. Second, by the classical reading the label's problem would seem to be its false positives, since the FDR is large and the FOR small. The audit in Section~\ref{sec:application} overturns this: 86 percent of the postings the label flags against the employer's field are remote by their own text, and the field is what is wrong. Third, the shrinkage of the labeled gap is larger than the contamination alone explains, which leads to the two ways a miss can bite.

\begin{table}[htbp]\centering\small\setlength{\tabcolsep}{4pt}
\caption{Where the language-model label errs, by propensity-score cell}\label{tab:errors}
\begin{tabular}{crrrrrrrr}\toprule
 & \multicolumn{2}{c}{Share remote} & \multicolumn{2}{c}{Classifier error} & \multicolumn{2}{c}{Column contamination} & \multicolumn{2}{c}{Pay gap} \\
\cmidrule(lr){2-3}\cmidrule(lr){4-5}\cmidrule(lr){6-7}\cmidrule(lr){8-9}
Cell & true $p(w)$ & label $\hat p(w)$ & FPR & miss rate & FDR & FOR & true $\tau(w)$ & label $\hat\tau(w)$ \\\midrule
1 & 0.017 & 0.007 & 0.003 & 0.73 & 0.38 & 0.012 & 0.417 & 0.141 \\
2 & 0.040 & 0.012 & 0.002 & 0.75 & 0.19 & 0.030 & 0.297 & 0.150 \\
3 & 0.037 & 0.019 & 0.003 & 0.58 & 0.15 & 0.022 & 0.236 & 0.146 \\
4 & 0.068 & 0.035 & 0.006 & 0.57 & 0.17 & 0.040 & 0.259 & 0.191 \\
5 & 0.098 & 0.050 & 0.008 & 0.57 & 0.15 & 0.059 & 0.136 & 0.128 \\
6 & 0.122 & 0.057 & 0.009 & 0.60 & 0.14 & 0.078 & 0.186 & 0.117 \\
7 & 0.142 & 0.075 & 0.010 & 0.53 & 0.11 & 0.081 & 0.177 & 0.164 \\
8 & 0.186 & 0.097 & 0.020 & 0.57 & 0.17 & 0.117 & 0.128 & 0.082 \\
9 & 0.240 & 0.136 & 0.021 & 0.50 & 0.12 & 0.139 & 0.157 & 0.127 \\
10 & 0.394 & 0.199 & 0.023 & 0.53 & 0.07 & 0.261 & 0.050 & -0.036 \\
\bottomrule\end{tabular}
\par\smallskip\parbox{0.95\linewidth}{\footnotesize Notes: Qwen label against the employer's remote field; cells are deciles of the label's linear propensity score (industry, experience level, work type), about 3,560 postings each. FPR = share of non-remote postings flagged; miss rate = share of remote postings not flagged; FDR = share of flagged postings that are not remote; FOR = share of unflagged postings that are remote. $\tau(w)$ is the within-cell gap in log pay between remote and non-remote postings; $\hat\tau(w)$ the gap between flagged and unflagged.}
\end{table}

\paragraph{Attenuation within a cell: notation.} Everything in this subsection is conditional on a cell, and $w$ always denotes a cell, that is, one value of the cell variable $W$ (in the application, one of the ten propensity-score deciles; with industry cells it would be one industry). Within a cell there are four groups of units, defined by their true status $d\in\{0,1\}$ and their label $\ell\in\{0,1\}$, and two kinds of mean outcome.
\begin{itemize}
\item $\mu_d(w)=\E[Y\mid D=d,\,W=w]$ is the mean outcome of the units in cell $w$ whose true status is $d$, whatever their label. Under Assumption~\ref{as:ign2} it equals the mean potential outcome $\E[Y(d)\mid W=w]$, so $\tau(w)=\mu_1(w)-\mu_0(w)$ is the true effect in the cell.
\item $\mu_{d\ell}(w)=\E[Y\mid D=d,\,\Dh=\ell,\,W=w]$ is the mean outcome of the units in cell $w$ with true status $d$ \emph{and} label $\ell$. There are four: $\mu_{11}$ for treated units the label catches, $\mu_{10}$ for treated units it misses, $\mu_{01}$ for its false positives, and $\mu_{00}$ for untreated units it correctly leaves unflagged.
\item $\Delta_1(w)=\mu_{11}(w)-\mu_{10}(w)$ is how much more the caught treated units earn than the missed ones, and $\Delta_0(w)=\mu_{01}(w)-\mu_{00}(w)$ how much more the false positives earn than the correctly unflagged untreated units. They measure whether the label's errors are \emph{selective}. I use this word throughout for errors that are related to the outcome: the label's errors are selective when the units it gets wrong differ in the outcome from the units of the same true status that it gets right, so that $\Delta_1\neq0$ or $\Delta_0\neq0$. The classical misclassification literature calls the opposite case non-differential error; selective errors are differential errors, named for what they do.
\end{itemize}
The regression sees only the two column means, $\E[Y\mid\Dh=1,w]$ and $\E[Y\mid\Dh=0,w]$, and their gap $\hat\tau(w)$.

\begin{lemma}[The labeled gap in a cell]\label{lem:gap}
In each cell $w$,
\begin{equation}\label{eq:gap}
\hat\tau(w)=\underbrace{\big\{1-\FDR(w)-\FOR(w)\big\}\,\tau(w)}_{\text{contamination (Aigner)}}\;+\;\underbrace{\kappa_1(w)\,\Delta_1(w)}_{\text{selective misses}}\;+\;\underbrace{\kappa_0(w)\,\Delta_0(w)}_{\text{selective false positives}},
\end{equation}
with non-negative weights
\begin{equation}\label{eq:kappa}
\begin{aligned}
\kappa_1(w)&=\{1-\FDR(w)\}\,m(w)+\FOR(w)\{1-m(w)\},\\
\kappa_0(w)&=\FDR(w)\{1-a(w)\}+\{1-\FOR(w)\}\,a(w).
\end{aligned}
\end{equation}
The identity is exact; Assumption~\ref{as:ign2} is needed only to read $\tau(w)$ as a causal effect. If the label's errors are not selective, $\Delta_1(w)=\Delta_0(w)=0$ and $\hat\tau(w)=\{1-\FDR(w)-\FOR(w)\}\tau(w)$.
\end{lemma}

Appendix~\ref{app:proofs} derives~\eqref{eq:gap} step by step. The idea is simple. Each column mean is a mixture: the flagged column holds a share $1-\FDR$ of treated units, whose mean is $\mu_{11}$, and a share $\FDR$ of untreated units, whose mean is $\mu_{01}$; the unflagged column holds a share $\FOR$ of treated units with mean $\mu_{10}$ and a share $1-\FOR$ of untreated units with mean $\mu_{00}$. If the label's errors were unrelated to the outcome, each of these four means would equal the corresponding group mean, $\mu_1$ or $\mu_0$, and the gap between the two mixtures would be the true gap times one minus the two contamination shares. That is the first term. In general the four means differ from the group means, and the second and third terms collect the differences.

\paragraph{What is Aigner's and what is new.} \citet{Aigner1973Binary} studies the regression of an outcome on a misclassified binary regressor with no controls, a common effect, and errors that are independent of the outcome given the true regressor (``non-differential'' errors). Under those assumptions the coefficient on the label is the true coefficient times an attenuation factor that, written in terms of the columns rather than the classifier, is $1-\FDR-\FOR$. The first term of~\eqref{eq:gap} is this result applied within each cell, where the effect $\tau(w)$ may differ from cell to cell; it is not new, and I write it in the contamination form because that form shows why the minority is hit hardest (equation~\eqref{eq:fdrfor}). The general identity~\eqref{eq:gap} drops the non-differential assumption. It is elementary algebra, but writing it this way makes two things visible that matter for what follows: what the extra terms are (the outcome gaps $\Delta_1$, $\Delta_0$ between the units the label gets wrong and the units it gets right), and how much each counts (the weights $\kappa_1$, $\kappa_0$).

\paragraph{Why a language model's errors can be selective.} A classical misclassification model treats the label as a noisy reading of the truth: a coin is flipped for each unit, with probabilities that depend on the unit's true status and perhaps its cell, and nothing else. Then which treated units are missed is random, $\Delta_1=0$, and likewise $\Delta_0=0$. A language model does not flip coins. It reads the posting, and the posting is also where the employer describes the job, its seniority, its duties and often its pay. Whatever makes a posting harder to read correctly may also be related to what the job pays. In the LinkedIn data the model reliably catches remote postings that say ``this is a remote position'' near the top, and misses remote postings whose cue sits in the middle or at the end of a long description, or in a benefits paragraph: the postings it misses are about twice as long as those it catches. Those postings also pay more. So among remote postings in the same cell, the missed ones pay more than the caught ones: $\Delta_1(w)<0$. The error rates $a(w)$, $m(w)$ and the contaminations $\FDR(w)$, $\FOR(w)$ cannot reveal this, because they count \emph{how many} units the label gets wrong; selectivity is about \emph{which} units, and is measured by the outcome gaps $\Delta_1$ and $\Delta_0$. Existing corrections for machine-generated regressors assume it away: \citet[Assumption~3(iii)]{Christensen2026Bootstrap} require the label's errors to be uncorrelated with the regression error up to terms smaller than the sampling error, which rules out selective errors of the size found here.

\paragraph{Reading the lemma: the misses carry a large weight.} The weights in~\eqref{eq:kappa} answer the question of how much each kind of error matters. In the contamination term the misses enter only through $\FOR(w)$, which is small whenever the treatment is rare (Table~\ref{tab:errors}); a reader who uses only that term, the classical formula, would conclude that misses are nearly harmless and that the damage comes from the false positives through $\FDR(w)$. The selective-miss term says otherwise. Its weight is $\kappa_1(w)\approx m(w)$ when the FDR and FOR are small: the outcome gap between caught and missed treated units is weighted by the miss \emph{rate}, the share of the treated who are missed, not by the share of the unflagged column they occupy. The reason is that misses damage the regression twice: they remove high-paying treated units from the flagged column, lowering its mean by $m\,\Delta_1$, and they place them in the unflagged column, raising its mean. With a miss rate of a half, half of any pay difference between caught and missed treated units passes into the labeled gap. The false positives' selectivity is weighted by $\kappa_0(w)\approx\FDR(w)$, the share of the flagged column they occupy. One practical consequence: a correction that divides the labeled gap by $1-\FDR-\FOR$, the classical fix, inverts only the first term and is wrong whenever the errors are selective; the correction of Section~\ref{sec:procedure} uses no error rates and is not affected.

Table~\ref{tab:delta} carries out the decomposition for the language-model label against the employer field. The weight on selective misses, $\kappa_1$, is between 0.45 and 0.62 in every cell. The missed remote postings pay more than the caught ones in nine of the ten cells, by up to 0.31 log points, so the selective-miss term is negative and as large as $-0.14$. The false positives, conversely, pay more than the correctly unflagged non-remote postings ($\Delta_0>0$ in every cell), which is what one would expect if many of them are in fact remote, as the audit confirms; the selective-false-positive term is positive and small, and offsets part of the contamination term's penalty on the FDR. In the cell where remote is rarest, the true gap of 0.42 becomes 0.25 after contamination, 0.11 after the selective misses, and 0.14 after the small offset from the false positives, which is the labeled gap.

\begin{table}[htbp]\centering\small
\caption{Decomposing the labeled pay gap in each cell (Lemma~\ref{lem:gap})}\label{tab:delta}
\resizebox{\linewidth}{!}{%
\begin{tabular}{crrrrrrrrr}\toprule
 & & Aigner term & \multicolumn{3}{c}{Selective misses} & \multicolumn{3}{c}{Selective false positives} & \\
\cmidrule(lr){4-6}\cmidrule(lr){7-9}
Cell & $\tau(w)$ & $(1-\FDR-\FOR)\tau$ & $\kappa_1$ & $\Delta_1$ & $\kappa_1\Delta_1$ & $\kappa_0$ & $\Delta_0$ & $\kappa_0\Delta_0$ & $\hat\tau(w)$ \\\midrule
1 & 0.417 & 0.251 & 0.45 & -0.306 & -0.139 & 0.39 & +0.075 & +0.029 & 0.141 \\
2 & 0.297 & 0.233 & 0.62 & -0.141 & -0.087 & 0.19 & +0.023 & +0.004 & 0.150 \\
3 & 0.236 & 0.195 & 0.50 & -0.152 & -0.076 & 0.15 & +0.177 & +0.027 & 0.146 \\
4 & 0.259 & 0.206 & 0.49 & -0.142 & -0.070 & 0.17 & +0.317 & +0.054 & 0.191 \\
5 & 0.136 & 0.107 & 0.51 & -0.018 & -0.009 & 0.16 & +0.187 & +0.030 & 0.128 \\
6 & 0.186 & 0.146 & 0.55 & -0.096 & -0.052 & 0.14 & +0.160 & +0.023 & 0.117 \\
7 & 0.177 & 0.143 & 0.51 & +0.013 & +0.007 & 0.12 & +0.118 & +0.014 & 0.164 \\
8 & 0.128 & 0.092 & 0.52 & -0.094 & -0.049 & 0.18 & +0.219 & +0.040 & 0.082 \\
9 & 0.157 & 0.117 & 0.51 & -0.023 & -0.012 & 0.14 & +0.164 & +0.022 & 0.127 \\
10 & 0.050 & 0.033 & 0.61 & -0.115 & -0.071 & 0.09 & +0.015 & +0.001 & -0.036 \\
\bottomrule\end{tabular}}
\par\smallskip\parbox{0.95\linewidth}{\footnotesize Notes: log points; Qwen label against the employer's remote field; same cells as Table~\ref{tab:errors}. $\Delta_1$ = mean log pay of remote postings the label catches minus those it misses; $\Delta_0$ = mean log pay of non-remote postings it flags minus those it correctly leaves unflagged. $\kappa_1$ and $\kappa_0$ are the weights in~\eqref{eq:gap}. In every row the Aigner term plus $\kappa_1\Delta_1+\kappa_0\Delta_0$ equals $\hat\tau(w)$ exactly.}
\end{table}

\subsection{Putting the two results together: two channels}
\label{sec:channels}

\citeauthor{Sloczynski2022OLS}'s decomposition~\eqref{eq:sloczynski} holds for any binary regressor, so it holds for the label: $\hat\beta=\hat w_1\hat\tau_1+\hat w_0\hat\tau_0$, with every quantity computed on the flagged and unflagged columns in place of the treated and untreated.

\begin{proposition}[Two channels]\label{prop:channels}
Let $\hat\beta$ be the coefficient on $\Dh$ in the regression with cell dummies. Then
\begin{equation}\label{eq:channels}
\hat\beta-\beta=\underbrace{\hat w_1(\hat\tau_1-\tau_1)+\hat w_0(\hat\tau_0-\tau_0)}_{\text{effects channel}}+\underbrace{(\hat w_1-w_1)(\tau_1-\tau_0)}_{\text{weights channel}}.
\end{equation}
\end{proposition}

The label damages both parts of the weighted average. Through the \emph{effects channel}, each group effect is replaced by its attenuated, and under selective errors further shifted, labeled version~\eqref{eq:gap}. Through the \emph{weights channel}, the label changes who is in which group, and therefore the weights. A label that misses half of the treated units makes the treated group look half as large: in the application the flagged share is 6.9 percent against a true remote share of 13.4. By~\eqref{eq:sloczynski} a smaller minority gets a larger weight, so $\hat w_1$ rises from $0.81$ to $0.92$. The regression now leans even harder on the treated group's effect, which is the effect the label has just attenuated most. The two channels compound: the label shrinks the minority's effect and then tells OLS to trust it more. With the language-model label, $\hat\tau_1=0.076$ against $\tau_1=0.127$, $\hat\tau_0=0.192$ against $0.248$, and $\hat\beta=0.086$ against $\beta=0.150$ (Table~\ref{tab:linkedin}); the weights channel is small in absolute terms here, because the two group effects are not far apart, but it has the same sign as the effects channel and grows with the gap between the ATT and the ATU.

\paragraph{A familiar decomposition.} Equation~\eqref{eq:channels} is the decomposition of \citet{Kitagawa1955Components} for a difference between two weighted averages, the same algebra as the Oaxaca--Blinder decomposition of a wage gap \citep{Oaxaca1973Male,Blinder1973Wage}. Each coefficient is a weighted average of two group effects, $\beta=w_1\tau_1+w_0\tau_0$ and $\hat\beta=\hat w_1\hat\tau_1+\hat w_0\hat\tau_0$. Their difference splits into a part that changes the effects at fixed weights, the counterpart of the Oaxaca--Blinder ``coefficients'' or unexplained component, and a part that changes the weights at fixed effects, the counterpart of the composition or ``endowments'' component. The weights channel is a single term because the weights sum to one: weight moved onto the treated is taken from the untreated, so a change in the mix moves the average only through the gap $\tau_1-\tau_0$. The link is not a coincidence. \citet{Kline2011Oaxaca} and \citet{Sloczynski2015Oaxaca} show that the Oaxaca--Blinder unexplained component is itself a treatment-effect estimator, and the weights of \citet{Sloczynski2022OLS} that I use throughout are closely related to that result. What is specific here is that the two weighted averages do not describe two populations. They describe one estimator, computed once with the true treatment and once with the label, so the decomposition attributes the label's bias to what the label does to the effects and what it does to the weights.

As with any decomposition of this kind, the split depends on the reference point. Evaluating the effects channel at the true weights and the weights channel at the labeled effects gives the equally valid
\begin{equation*}
\hat\beta-\beta=w_1(\hat\tau_1-\tau_1)+w_0(\hat\tau_0-\tau_0)+(\hat w_1-w_1)(\hat\tau_1-\hat\tau_0),
\end{equation*}
and the two versions differ by the interaction $(\hat w_1-w_1)\{(\hat\tau_1-\tau_1)-(\hat\tau_0-\tau_0)\}$. In the application the choice does not matter. With the employer field as the truth and the language-model label, the effects channel is $-0.051$ and the weights channel $-0.013$ in the ordering of~\eqref{eq:channels}, against $-0.052$ and $-0.013$ in the reverse ordering, and the interaction is $0.0005$. About 80 percent of the bias runs through the effects and 20 percent through the weights whichever way it is computed, because the label attenuates the two group effects by similar amounts ($-0.051$ and $-0.056$).

A useful special case: when the treatment is rare and the label misses little, $\hat w_1\approx1$ and $\hat\beta\approx(1-\FDRbar)\,\ATT$, the effect on the treated shrunk by the pooled false discovery rate of the flagged column. The regression on a label is then not about the population at all; it is about the flagged units, discounted for the share of them that are not treated. None of this depends on the sample size.

\subsection{What the audit must measure: recall, cell by cell, and which units are missed}
\label{sec:measure}

The two results of the previous subsections point to the same place. OLS reports mostly the effect on the minority (Section~\ref{sec:ols}); that effect is read from the gap between the treated units and the untreated units at the same propensity scores; and a label damages that gap most through the treated units it fails to flag, which sit in the unflagged column exactly where the treated are, and through the weight it moves onto the damaged effect. So the quantity a researcher has to know about the label is not its precision but its \emph{recall}: the share of truly treated units it catches, cell by cell, and which ones it misses. This subsection shows, for the language-model label, why a precision check passes, what the recall check finds, and how the misses are related to the outcome.

\paragraph{Precision passes.} Precision is the share of flagged units that are truly treated, one minus the FDR. Against the posting text it is about 95 percent for the language-model label (Section~\ref{sec:application}); even against the employer's field it is above 80 percent in eight of the ten cells (Table~\ref{tab:errors}), and the false-positive rate is below one percent everywhere. A researcher who read a sample of flagged postings would have found a label that is almost always right when it says ``remote'', and would have stopped there. Table~\ref{tab:whicherror} below shows what that check misses: removing the label's false positives changes the coefficient by less than a hundredth of a log point.

\paragraph{Recall: how many treated units the label misses, and which.} The quantity that matters is the label's \emph{recall}, the share of truly treated units it catches, one minus the miss rate. Precision, the share of flagged units that are truly treated, is what validations usually report; recall is what the regression suffers from. And recall has two sides: how many treated units are missed, and which ones. Table~\ref{tab:recall} gives both for the language-model label.

\begin{table}[htbp]\centering\small
\caption{Recall: how many remote postings the label misses, and which}\label{tab:recall}
\resizebox{\linewidth}{!}{%
\begin{tabular}{crrrr@{\hskip 18pt}crr@{\hskip 18pt}crr}\toprule
\multicolumn{5}{c}{A. By propensity cell} & \multicolumn{3}{c}{B. By pay, within cell} & \multicolumn{3}{c}{C. By length, within cell}\\
\cmidrule(r{12pt}){1-5}\cmidrule(r{12pt}){6-8}\cmidrule{9-11}
Cell & Remote & Missed & Miss rate & Share & Pay quintile & Miss rate & Pay & Length quintile & Miss rate & Pay\\\midrule
1 & 60 & 44 & 0.73 & 0.02 & 1 & 0.50 & -0.73 & 1 & 0.45 & -0.05 \\
2 & 142 & 107 & 0.75 & 0.04 & 2 & 0.55 & -0.23 & 2 & 0.50 & -0.03 \\
3 & 133 & 77 & 0.58 & 0.03 & 3 & 0.53 & +0.05 & 3 & 0.55 & -0.03 \\
4 & 242 & 137 & 0.57 & 0.05 & 4 & 0.58 & +0.28 & 4 & 0.62 & +0.01 \\
5 & 350 & 199 & 0.57 & 0.08 & 5 & 0.59 & +0.64 & 5 & 0.63 & +0.09 \\
6 & 436 & 261 & 0.60 & 0.10 & & & & & & \\
7 & 504 & 266 & 0.53 & 0.10 & & & & & & \\
8 & 661 & 375 & 0.57 & 0.14 & & & & & & \\
9 & 856 & 428 & 0.50 & 0.16 & & & & & & \\
10 & 1,403 & 743 & 0.53 & 0.28 & & & & & & \\
\midrule All & 4,787 & 2,637 & 0.55 & 1.00 & & & & & & \\
\bottomrule\end{tabular}}
\par\smallskip\parbox{0.97\linewidth}{\footnotesize Notes: remote postings by the employer's field; ``missed'' = not flagged by the language model. Panel A: by the ten cells of Table~\ref{tab:errors}; ``share'' = the cell's share of all missed remote postings. Panels B and C: remote postings are ranked into quintiles within their cell, by log pay (B) or by posting length in tokens (C); ``pay'' is the mean log pay relative to the cell mean of remote postings, so cell composition plays no role.}
\end{table}

\emph{How many, and where.} Panel A shows that the miss \emph{rate} is not higher where remote work is common; if anything it is lower, from 73 to 75 percent in the two lowest cells to 50 to 53 percent in the two highest. What is higher at high propensity is the \emph{number} of misses, because that is where the remote postings are: the four highest cells hold 69 percent of all missed remote postings, and the top cell alone 28 percent. The count matters for the regression because of where it lands. A miss in a high-propensity cell is placed among the counterfactuals of the treated units that sit in the same cell, and those are the cells that carry most of the weight in the ATT and in $\beta$; there the contamination of the unflagged column, the FOR of Table~\ref{tab:errors}, reaches 26 percent.

\emph{Which ones.} Does that make the misses higher-paid? Not for the reason that the high-propensity cells pay more. The regression compares flagged and unflagged postings within cells, so differences in pay across cells do not enter the bias. Indeed the pooled pay gap between missed and caught remote postings, 6.3 log points, is entirely a within-cell gap: comparing each missed posting with the caught remote postings of its own cell gives 8.2 log points, and the between-cell part is $-1.9$, because the miss rate is slightly higher in the low-paying, low-propensity cells. Panels B and C show what lies behind the within-cell gap. Ranking remote postings within their cell by pay, the miss rate rises from 50 percent in the bottom fifth to 59 percent in the top fifth. Ranking them by length, it rises from 45 percent among the shortest postings (about 400 tokens) to 63 percent among the longest (about 1,500), and the longest postings pay 9 log points more than the cell average while the shortest pay 5 log points less. Holding the cell fixed, posting length predicts a miss more strongly than pay does (a gap of 18 percentage points between the longest and the shortest fifth, against 7 between the best- and worst-paid fifth when both enter together). The mechanism is the one the audit found reading the postings: the model catches remote jobs that announce it at the top of a short posting and misses those that mention it once in a long one, and long postings are for better-paid jobs. This is what makes $\Delta_1<0$ in Table~\ref{tab:delta}.

The two sides of recall therefore work together. The number of misses is largest where the treated are, so the counterfactual of the treated is the most contaminated; and within those cells the label misses the better-paid treated, so the contamination is selective and pushes the treated effect down. How much less harm would a label with the same recall do if its misses were random? The last row of Table~\ref{tab:whicherror} answers it: keeping the label's number of misses in every cell but reassigning them at random among the cell's remote postings raises the coefficient from 0.086 to 0.128, against a truth of 0.150. About two thirds of the bias comes from \emph{which} remote postings the label misses, and one third from \emph{how many}. A label whose recall rose only in the low-propensity cells would barely help, because few of the treated are there.

\paragraph{Which error drives the bias in the regression.} Table~\ref{tab:whicherror} confirms the reading with two counterfactual labels, computed on the full sample. Removing the label's false positives (setting the label to zero wherever the employer field says non-remote) leaves the bias intact: $\beta$ moves from $0.086$ to $0.077$. Removing the misses (setting the label to one wherever the field says remote) removes the bias entirely: $\beta$ is $0.153$ against a truth of $0.150$, and the ATT, the ATU and the weights return to their true values. The misses are the whole of the bias here.

\begin{table}[htbp]\centering\small
\caption{Which error drives the bias: counterfactual labels}\label{tab:whicherror}
\begin{tabular}{lccccc}\toprule
Regressor & $\beta$ & ATT & ATU & $\rho$ & $w_1$ \\\midrule
True treatment (employer field) & 0.150 & 0.140 & 0.214 & 0.134 & 0.81 \\
Language-model label (both errors) & 0.086 & 0.081 & 0.124 & 0.069 & 0.92 \\
Label with its false positives removed (misses only) & 0.077 & 0.073 & 0.125 & 0.060 & 0.92 \\
Label with its misses removed (false positives only) & 0.153 & 0.142 & 0.210 & 0.143 & 0.80 \\
\addlinespace
Same misses per cell, but chosen at random & 0.128 & 0.124 & 0.180 & 0.069 & -- \\
\bottomrule\end{tabular}
\par\smallskip\parbox{0.92\linewidth}{\footnotesize Notes: log points; same ten cells as Table~\ref{tab:errors}; truth = the employer's remote field. ``Misses only'' sets the label to zero for every posting the field marks non-remote; ``false positives only'' sets it to one for every posting the field marks remote. The last row keeps the label's number of misses in every cell but reassigns them at random among the cell's remote postings (average over 100 draws), so recall is unchanged and only the selection of the misses differs. $\rho$ is the flagged share and $w_1$ the weight on the flagged group's effect in~\eqref{eq:sloczynski}.}
\end{table}

Two routes carry the misses' damage, and both are invisible to the classical formula. The first is the selective-miss term of~\eqref{eq:gap}, weighted by the miss rate rather than the FOR. The second is the weights channel of the next subsection: by halving the flagged share, the misses change which cells OLS trusts and how much weight it puts on the treated group's effect. A researcher who validated the label by its precision, or who corrected it with the classical formula, would have been told that a label with a small FOR is safe. It is not, and the lesson for the audit is that the unflagged column must be read, and read where the treated are.

\paragraph{When misses carry the bias, and when false positives do.} The finding that misses, not false positives, carry the bias is the paper's central empirical claim, and it is what justifies reading the unflagged column. But it comes from one label and one dataset, and the published applications assume the opposite: \citet{Battaglia2024Inference} correct for false positives, measured on flagged records only, and \citet{Christensen2026Bootstrap} set the misses equal to the false positives. A reader can fairly ask whether misses dominate here because of something special about LinkedIn or about this language model. This subsection answers that question.

The aim is to show three things. First, which error dominates is not a property of a dataset as a whole. It is set by a few numbers that describe any label in any cell: the base rate, the number of misses per false positive, and how far the missed units and the falsely flagged units are from their true groups in the outcome. Second, the classical intuition that false positives dominate when the treatment is rare is right only when errors are random, and a modest selectivity of misses reverses it. Third, these numbers can be estimated from a pilot that reads both columns, so a researcher can tell which case they are in before spending the rest of the audit.

The analysis supports the claim in two ways. It explains the LinkedIn result instead of only reporting it: misses dominate there because they are numerous, because they are selective, and because the false positives look like the treated. And it shows that the audit design does not depend on the claim holding everywhere. Where false positives dominate, the same pilot says so; the one step that cannot be skipped is reading the unflagged column once.

Lemma~\ref{lem:gap} gives the answer exactly. Write $\delta_1(w)=\Delta_1(w)/\tau(w)$ and $\delta_0(w)=\Delta_0(w)/\tau(w)$ for the two selectivity gaps in units of the effect, and $R(w)=p(w)m(w)/[\{1-p(w)\}a(w)]$ for the number of misses per false positive in the cell.

\begin{proposition}[Which error carries the bias]\label{prop:regime}
In each cell, $\hat\tau-\tau=b_M+b_F+b_{MF}$, where $b_M$ is the bias of the label with its false positives removed, $b_F$ the bias of the label with its misses removed, and $b_{MF}$ the interaction. With the argument $w$ dropped,
\begin{align}
\frac{b_M}{\tau}&=-\FOR_M\{1-(1-m)\delta_1\}+m\,\delta_1, & \FOR_M&=\frac{pm}{pm+1-p},\label{eq:bM}\\
\frac{b_F}{\tau}&=-\FDR_F\{1-(1-a)\delta_0\}+a\,\delta_0, & \FDR_F&=\frac{m}{m+R}.\label{eq:bF}
\end{align}
The flagged share is below the true share, $\hat p<p$, exactly when $R>1$.
\end{proposition}

Appendix~\ref{app:proofs} gives the proof. Each error does damage in two ways. The first is \emph{contamination}: the misplaced units dilute the column they land in, by $\FOR_M$ for misses and $\FDR_F$ for false positives, and the damage is scaled by how far they are from that column's true members, the terms in braces. The second is \emph{selection}: the units of the same true status that the label gets right are no longer representative of their group, and this part is weighted by the error \emph{rate}, $m$ for misses and $a$ for false positives. Three consequences follow.

First, with errors that are not selective ($\delta_1=\delta_0=0$) the classical picture holds. When the treatment is rare $\FOR_M\approx pm$ is small and $\FDR_F=m/(m+R)$ is not, so false positives carry the bias unless misses outnumber them many times over: at $p=3$ percent, misses carry half of it only when $R$ exceeds about 33. This is the case that the corrections of \citet{Battaglia2024Inference} and \citet{Christensen2026Bootstrap} are built for.

Second, the selection part of the misses is weighted by the miss rate, which does not shrink with $p$, while that of the false positives is weighted by the false-positive rate, which is small for any useful label. A modest selectivity of misses therefore overturns the classical picture. At $p=3$ percent and $m=0.5$, misses carry half the bias from $R\approx3.1$ when $\delta_1=-0.25$, and from $R\approx1.4$ when $\delta_1=-0.5$, close to the calibrated case $R=1$.

Third, false positives do harm only to the extent that they look like the untreated. As $\delta_0$ rises towards one, the braces in~\eqref{eq:bF} vanish and $b_F/\tau$ falls to $a(1-\FDR_F)$. False positives that are near-treated, such as hybrid jobs, or truly treated units that the benchmark itself misses, barely move the gap. The last line of the proposition gives the weights channel: a label with more misses than false positives shrinks the apparent treated group and, by~\eqref{eq:sloczynski}, raises the weight on its damaged effect.

\begin{figure}[htbp]\centering
\includegraphics[width=\linewidth]{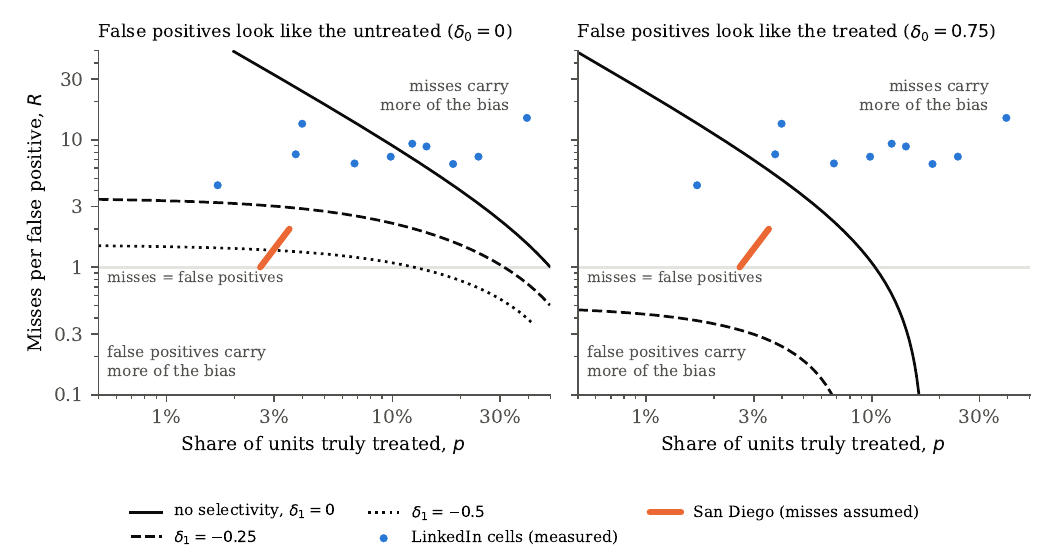}
\caption{Which error carries the bias: misses or false positives}\label{fig:regime}
\vspace{2pt}
\begin{minipage}{0.95\linewidth}\footnotesize\textit{Notes:} each curve is the boundary of Proposition~\ref{prop:regime} where misses and false positives contribute equally to a cell's bias ($|b_M|=|b_F|$), for a miss rate $m=0.5$ and three levels of miss selectivity $\delta_1$. Above and to the right of a curve, misses carry more of the bias. Left panel: false positives with the outcome of the untreated ($\delta_0=0$); right panel: false positives close to the treated ($\delta_0=0.75$), where the $\delta_1=-0.5$ boundary lies below the plotted range. Blue dots: the ten LinkedIn cells (Qwen label against the employer field). Orange: the San Diego application of \citet{Battaglia2024Inference}, with misses set equal to, or twice, the false positives, as in \citet{Christensen2026Bootstrap}; its miss rate was never measured.\end{minipage}
\end{figure}

Figure~\ref{fig:regime} draws the boundaries. The LinkedIn cells have 4 to 15 misses per false positive, a median $\delta_1$ of about $-0.5$, and $\delta_0$ between 0.08 and 1.7. In every cell misses carry between 75 and 99 percent of the bias. Weighted by treated postings, the miss part is $-0.064$ and the false-positive part $0.001$ of a total of $-0.059$, the same verdict as Table~\ref{tab:whicherror}. In the left panel several cells sit below the solid curve: had their misses been random and their false positives like the untreated, false positives would have carried more of their bias. San Diego, with roughly 2.6 to 3.5 percent of postings remote, is the opposite case on paper. With as many misses as false positives and random misses, false positives carry 97 percent of its bias. With $\delta_1=-0.5$ they carry 58 percent, and with twice as many misses as false positives, misses carry 58 percent. Which error matters there depends on two numbers, the miss count and the misses' selectivity, that only a read of the unflagged column can supply.

The proposition is also a diagnostic. A pilot that reads both columns estimates every input: $p$, $m$ and $a$ from the audited counts, $\delta_1$ and $\delta_0$ from the audited outcome means. If $\hat R$ is well above one or $\hat\delta_1$ is clearly negative, misses carry the bias and the second wave belongs in the unflagged column. If $\hat R$ is near one and the caught and missed treated units look alike, false positives carry it, and a precision check with a correction from error rates may be enough. Either way the label can be placed on the map only after its unflagged column has been read once.

\paragraph{Why misses are hard to measure, and what follows for the audit.}

The two errors are not equally easy to measure. False positives are cheap: the flagged column is small, and reading a sample of it gives $\FDR(w)$ directly. This is the check researchers usually run, under the name of precision, and it is what \citet{Battaglia2024Inference} do in their application, reading only the flagged records. Misses are expensive: they are a few percent of a column that holds nearly everything, so a random read of the unflagged column finds mostly true negatives and must be large before it pins down $\FOR(w)$. In the application a precision check passes the label with flying colours, 95 percent of its flags are correct by the text, while 55 percent of the remote postings sit unflagged. And the misses matter twice: they attenuate the treated group's effect by contaminating its counterfactual, and they shrink the apparent minority and tilt the weights.

Two features of the problem make the measurement feasible. First, the two contamination rates vary across cells in a known direction, so the audit can be stratified by cell and by column, reading the unflagged column hardest where the treated are most likely to hide (high propensity cells) and where a single miss stands for most records (the largest cells). Second, the correction does not need the rates at all: it needs, for each cell, the audited difference between the label and the truth, weighted up by the sampling design, which is robust to selective errors because it never rescales. That is the procedure of the next section.

\subsection{The precision trap}
\label{sec:trap}

\begin{figure}[htbp]\centering
\includegraphics[width=0.85\linewidth]{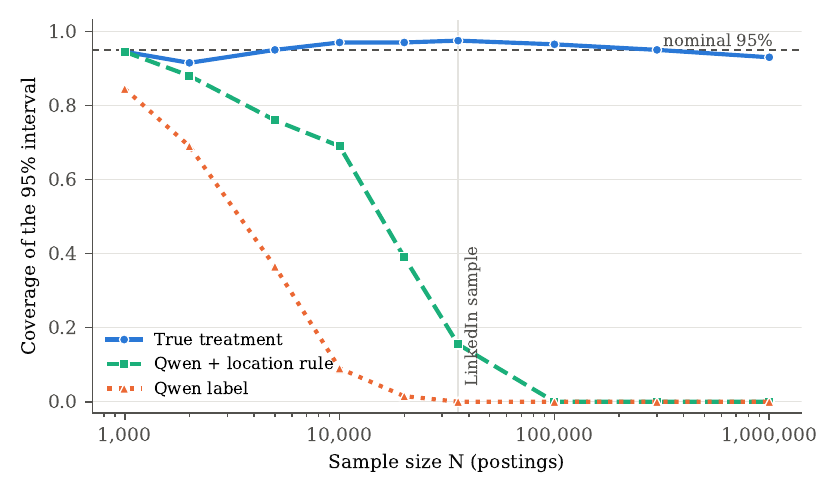}
\caption{The precision trap: coverage of the usual 95\% interval for $\beta$}\label{fig:coverage}
\par\smallskip\parbox{0.9\linewidth}{\footnotesize Notes: samples of size $N$ drawn with replacement from the 35,618 LinkedIn postings; the truth is the coefficient on the employer's remote field in the full sample, with cell dummies as controls; heteroskedasticity-robust intervals; 200 samples per point.}
\end{figure}

The bias $b=\hat\beta-\beta$ is a population quantity: it is set by the label's error rates and the outcome gaps, and labeling more units does not change it. The standard error, by contrast, falls like $\sigma/\sqrt N$. \citet{Battaglia2024Inference} draw the consequence for a generated regressor: the usual 95 percent interval around the OLS estimate covers the true $\beta$ with probability
\begin{equation}\label{eq:coverage}
\Phi(1.96-r_N)-\Phi(-1.96-r_N),\qquad r_N=\sqrt N\,b/\sigma,
\end{equation}
which falls to zero as $N$ grows. \citet{Christensen2026Bootstrap} formalize the same regime with local asymptotics in which the error rates shrink at the rate of the standard error, so that the bias and the sampling error stay of the same order; equation~\eqref{eq:coverage} is the fixed-error version. Language models make it cheap to label very large samples, so this is the regime that matters. Figure~\ref{fig:coverage} shows it in the application: with the true treatment the interval covers about 95 percent of the time at every sample size, but with the language-model label coverage is below 10 percent at 10,000 postings and zero at the size of the LinkedIn sample; even the better label, with the location rule of Section~\ref{sec:application}, covers 16 percent of the time at 35,618 postings and never beyond. The researcher ends up with a tight interval around a number that is biased by both channels and that, through the weights, describes an estimand nobody chose.

\section{The procedure: audit the label, correct the cells, choose the estimand}
\label{sec:procedure}

This section is the paper's practical contribution: a recipe a researcher can follow with an LLM-labeled treatment, the defaults I recommend, and what to report. Box~\ref{box:recipe} gives the recipe on one page; the rest of the section explains each step. The correction itself uses the logic of two-phase sampling and prediction-powered inference \citep{Neyman1934Two,ChenHongTamer2005Auxiliary,Angelopoulos2023PPI,Kluger2025Imputed}: the label gives a precise but biased estimate on the full sample, and a hand-labeled audit measures the bias, which is then subtracted. What is specific to this setting is \emph{where} the bias must be measured (in cells, in both columns), \emph{how} the audit must be read (blind to the outcome), and \emph{which} estimand is reported at the end.

\paragraph{Why precision is not enough.} Section~\ref{sec:bias} gives the reason the audit has to be designed this way, and it is worth restating in one paragraph. A label can fail in two ways, and a regression on the label suffers from both, but in different places. False positives contaminate the flagged column; they are easy to measure, because the flagged column is small and reading a sample of it gives its contamination directly. Misses contaminate the unflagged column; they are hard to measure, because they are a few percent of a column that holds most of the sample, and a random read finds mostly correctly unflagged units. Yet the misses are what matters most for the slope. OLS estimates mostly the effect on the treated; that effect is read from the gap between the treated and the untreated units that look like them; and the misses both remove treated units from the first side of that gap and add them to the second, selectively when the label reads the same text as the outcome, while also shrinking the apparent treated group and raising its weight. In the application, removing the label's misses removes the whole bias; removing its false positives removes none of it (Table~\ref{tab:whicherror}). The audit therefore samples both columns, in every cell, and aims the unflagged reads where treated units are likely to hide. A simple random validation sample, the design in \citet{LudwigMullainathanRambachan2026LLM}, does neither: with 13 percent of postings treated and 55 percent of those missed, a random sample of 250 postings contains about 18 misses, spread over ten cells, and the precision of the corrected estimate is governed by exactly those few records.

\begin{center}
\setlength{\fboxsep}{10pt}\fbox{\parbox{0.93\linewidth}{\small
\textbf{Box \refstepcounter{boxc}\theboxc\label{box:recipe}. The recipe.}\par\smallskip
\textbf{0. Define the treatment} as one written rule, including the ambiguous cases (e.g.\ hybrid work, remote offered as an option, field jobs). Look for structured fields that already carry the treatment and check them on a few records before trusting the text alone.\par
\textbf{1. Audit the label.}
(a) Cells: deciles of the label's linear propensity score $\hat p(X)$. Strata: cell $\times$ flagged/unflagged, optionally $\times$ any variable observed for every unit. Audit \emph{both} columns.
(b) Pilot: 150--300 records, proportional, at least 2 per stratum.
(c) Read blind: outcome text masked, no outcome field, no label, no stratum, records shuffled. Two readers (a language model with a fixed protocol and a person) code each record independently with a verbatim quote; reconcile disagreements with one written rule. Silent records are coded \emph{undetermined}, never resolved with the outcome.
(d) Diagnose: from the pilot, estimate the misses per false positive $R$ and the selectivity gaps $\delta_1,\delta_0$, and place the label on the map of Proposition~\ref{prop:regime}.
(e) Second wave by Neyman allocation, $n_s\propto N_sS_s$, with $S_s$ from the pilot and a rule-of-three floor where the pilot saw no error. Size it with the budget rule~\eqref{eq:budget}.\par
\textbf{2. Correct the cells.} For each cell, replace the label's cell moments by the corrected moments~\eqref{eq:corrected}: the full-sample label average plus the design-weighted audit gap.\par
\textbf{3. Choose the estimand.} Report ATT, ATU and the OLS coefficient from the corrected moments, together with \citeauthor{Sloczynski2022OLS}'s $\rho$ and $w_1$, so the reader sees whose effect the regression reports. Say which estimand answers the question.\par
\textbf{Report:} naive, corrected and audit-only estimates; two-phase bootstrap intervals; the audit design and sizes; reader agreement before reconciliation; the two bounds for undetermined records.\par
\textbf{Software:} the \texttt{llmaudit} Python package implements every step, including the blind sheets, the labeling tool and the language-model reader protocol.}}
\end{center}

\subsection{Step 1: audit the label}

\paragraph{Cells: the label's propensity score.} The cells are bins of the \emph{label's} propensity score, $\hat p(X)=\Pr(\Dh=1\mid X)$, the probability that the label flags a unit given the controls, estimated by a linear regression of the label on the control dummies and cut into deciles. Three things recommend it. It can be computed for every unit without knowing the truth, which the true propensity score cannot. It is the score that \citeauthor{Sloczynski2022OLS}'s decomposition uses for the labeled regression, so the cells are the right coordinate for the weights channel. And it orders the units by how likely the label is to flag them, which is also how the contamination rates~\eqref{eq:fdrfor} vary: the false discovery rate is highest in the low-score cells and the misses, in absolute number, pile up in the high-score cells, so the audit can be aimed. Within a cell, flagged and unflagged units are compared, so the cells are also the control set of the final regression; they coarsen the original controls, and I state the estimands as conditional on the cells.

\paragraph{Where to look.} The audit has to measure, in every cell, how the label's two columns differ from the true ones, so it is stratified by cell and by flag. Checking only flagged records (precision) is the common practice and it misses the bias that matters here: under heterogeneity the effect on the treated is learned from the untreated line at high propensity scores, which is exactly where the missed treated records sit. In the application the label is about 95 percent precise against the posting text, yet the naive coefficient is nearly 40 percent too low, because of what sits in the unflagged column. The pilot then says where the second wave should go: Proposition~\ref{prop:regime} turns its estimates of the misses per false positive and of the selectivity gaps into the share of each cell's bias that comes from each column, and the second wave is spent on the column that carries it.

\paragraph{How to read.} The audit's purpose is to establish the truth, so the reader must not see the outcome: a reader who sees pay can be drawn to call high-paying postings remote, which builds the outcome into the truth it is supposed to test. I mask the outcome's amounts (not whole sentences, which can carry the treatment too) and drop the outcome field. I use two readers, a language model following a fixed protocol and a person, who code each record independently and quote the deciding phrase; their disagreements are resolved by a written rule, not case by case. When the text is silent the record is \emph{undetermined}; I report the corrected estimates under two assignments (follow a structured field where one exists, or not treated), which bound the answer. When the person cannot read every record, the person reads all records the model reader codes as treated or ambiguous, all records where a structured field disagrees with it, and a random, weight-proportional sample of the rest; the share of model calls the person overturns in that sample corrects the model reader on the records nobody else read, by the same two-phase logic.

\paragraph{How much to read.} After a proportional pilot, the second wave allocates by Neyman's rule, $n_s\propto N_sS_s$, where $N_s$ is the stratum size and $S_s$ the spread of the correction term $\{D_i-\Dh_i\}\{Y_i-\bar\mu_1\}$ in the pilot. In a stratum where the pilot found no error I set $S_s$ from the rule-of-three bound $3/n_s$ on the error rate rather than to zero; otherwise the formula would starve exactly the strata whose error rate is unknown. Waves drawn with different designs are combined with inclusion probabilities $\pi_i=1-\prod_j(1-\pi_{ij})$.

\paragraph{The pilot diagnostic in practice.} The \texttt{llmaudit} package implements Proposition~\ref{prop:regime} as a pilot diagnostic. From the audited records of both columns it estimates each cell's base rate, error rates and selectivity gaps, splits the cell's bias into its miss and false-positive parts, and says where the second wave should go (Table~\ref{tab:pilotdiag}). Two choices make it usable on a pilot. A pilot sees only a few misses in each cell, and noise in a cell's selectivity gap always inflates the miss part, so the diagnostic pools the gaps within cells across cells. And it gives a verdict only when its bootstrap interval allows one.

On the LinkedIn audits, the first 350 postings already settle the question for the Qwen label. Misses carry 91 percent of its bias (90 percent interval 0.58 to 0.99), with about 22 misses per false positive, so the second wave belongs in the unflagged column; the full 650 postings change nothing. For the label with the location rule, which recovers many of the misses, the share falls to about one half and the diagnostic sends the second wave to both columns, as my Neyman wave did. In synthetic data with a known answer (panel B), a 400-record pilot recovers a miss share of 0.93 as 0.91. When errors are random and false positives plentiful, a 400-record pilot that sees only four misses declines to give a verdict, and a 1,500-record pilot sends the second wave to the flagged column.

\begin{table}[htbp]\centering\small
\caption{The pilot diagnostic: which error carries the bias, and where the second wave goes}\label{tab:pilotdiag}
\resizebox{\linewidth}{!}{\begin{tabular}{lrrrrcl}\toprule
 & Audited & Misses & False & Misses per & Share of bias from misses & Second wave \\
 & records & audited & positives & false positive & [90\% interval] & \\ \midrule
\multicolumn{7}{l}{\textit{A. LinkedIn audits (truth: blind reconciled reading)}}\\
Qwen, pilot & 350 & 62 & 8 & 22.0 & 0.91 [0.58, 0.99] & unflagged column \\
Qwen, all audits & 650 & 73 & 14 & 21.3 & 0.91 [0.57, 0.99] & unflagged column \\
Qwen + location rule (L2), pilot & 350 & 28 & 12 & 2.7 & 0.51 [0.13, 0.84] & both columns (Neyman) \\
Qwen + location rule (L2), all audits & 650 & 39 & 18 & 4.1 & 0.53 [0.08, 0.89] & both columns (Neyman) \\
\addlinespace\multicolumn{7}{l}{\textit{B. Synthetic checks (true share of bias from misses in parentheses)}}\\
selective misses, 1\% false positives & 400 & 23 & 16 & 7.7 & 0.91 [0.81, 0.97] (0.93) & unflagged column \\
random errors: 10\% misses, 4\% false positives & 400 & 4 & 25 & 0.4 & 0.45 [0.12, 0.80] (0.12) & both columns (Neyman) \\
random errors: 10\% misses, 4\% false positives & 1,500 & 16 & 106 & 0.4 & 0.21 [0.02, 0.52] (0.12) & flagged column \\
\bottomrule\end{tabular}}
\par\smallskip\begin{minipage}{0.97\linewidth}\footnotesize\textit{Notes:} output of \texttt{llmaudit.pilot\_diagnostic}. Shares are treated-weighted averages of the cell-level split of Proposition~\ref{prop:regime}, with the selectivity gaps pooled within cells; intervals from 300 bootstrap draws within audit strata. Panel A: Qwen label and the label with the location rule of Section~\ref{sec:application}, silent postings coded not remote; the pilot is the first 350 audited postings, read before the Neyman wave. Against the blind reading most of the label's apparent false positives are remote, so misses per false positive exceed the figures against the employer field in Figure~\ref{fig:regime}. Panel B: the synthetic design of the package tests (60,000 units, 30 groups); proportional pilots with a reserved flagged share. A verdict requires the interval to clear one half, or the point estimate to lean past 0.6 (0.4) with the interval short of the opposite lean; otherwise the diagnostic sends the second wave to both columns.\end{minipage}
\end{table}

\subsection{Step 2: correct the cells}

This step is standard, and I use it as it stands. It is a stratified difference estimator from two-phase sampling and prediction-powered inference \citep{Neyman1934Two,ChenHongTamer2005Auxiliary,Angelopoulos2023PPI,Fisch2024StratPPI,Kluger2025Imputed}; \citet[Appendix~C.2]{LudwigMullainathanRambachan2026LLM} give it for a regression with an LLM label as a covariate, and \citet{Barnatchez2025TwoPhase} give efficient doubly robust estimators of the average treatment effect when the treatment and the outcome are mismeasured and validated on a second-phase sample. What is specific to this use is only the choice of what to correct: the cell moments rather than the regression coefficient, so that the ATT, the ATU, the coefficient and \citeauthor{Sloczynski2022OLS}'s decomposition all come from the same corrected inputs. Nothing in the audit design of Step~1 or the choice of estimand in Step~3 depends on this particular estimator; a researcher can substitute an efficient estimator such as that of \citet{Barnatchez2025TwoPhase} and still report the ATT, the ATU and the weights. I use corrected cell means because \citeauthor{Sloczynski2022OLS}'s decomposition is a function of them and because they are transparent.

Let $A$ be the audited set, $\pi_i$ the inclusion probabilities, and $N_w$ the number of units in cell $w$. For each cell,
\begin{equation}\label{eq:corrected}
\begin{aligned}
\tilde m_{YD}(w)&=\frac1{N_w}\sum_{i\in w}Y_i\Dh_i+\frac1{N_w}\sum_{i\in A\cap w}\frac{Y_i(D_i-\Dh_i)}{\pi_i},\\
\tilde m_D(w)&=\frac1{N_w}\sum_{i\in w}\Dh_i+\frac1{N_w}\sum_{i\in A\cap w}\frac{D_i-\Dh_i}{\pi_i}.
\end{aligned}
\end{equation}
The first term is the label's cell moment on the full sample; the second is the audit's estimate of how far it is from the truth. The corrected treated share, the two cell means and the cell effect follow: $\tilde p(w)=\tilde m_D(w)$, $\tilde\mu_1(w)=\tilde m_{YD}(w)/\tilde p(w)$, $\tilde\mu_0(w)=\{\bar Y(w)-\tilde m_{YD}(w)\}/\{1-\tilde p(w)\}$ and $\tilde\tau(w)=\tilde\mu_1(w)-\tilde\mu_0(w)$, where $\bar Y(w)$ is the cell's mean outcome, which the label does not affect.

The properties of this estimator are those of any stratified two-phase estimator; I state them for completeness, because the last of them is what sizes the audit.

\begin{proposition}[Corrected estimates and the audit floor]\label{prop:audit}
Suppose every stratum is audited with known $\pi_i>0$. Then: (i) conditional on the full sample, the corrected moments~\eqref{eq:corrected} are unbiased for the true cell moments, whatever the label's errors and whether or not they depend on the outcome; (ii) the corrected ATT, ATU, ATE and $\beta$, which are smooth functions of these moments, are consistent as the audit grows in every stratum; (iii) their variance is approximately
\begin{equation}\label{eq:variance}
\Var(\tilde\theta)\approx\frac{\sigma_1^2}{N}+\sum_s W_s^2\Big(1-\frac{n_s}{N_s}\Big)\frac{S_{\psi,s}^2}{n_s},
\end{equation}
where $W_s=N_s/N$ and $S_{\psi,s}$ is the within-stratum spread of the estimand's influence through the label error $D_i-\Dh_i$. The second term does not shrink with $N$: for a fixed audit, precision stops improving however many records the language model labels. Under Neyman allocation the audit needed to bring the second term below a target variance $v$ is
\begin{equation}\label{eq:budget}
n^\ast=\Big(\sum_s W_sS_{\psi,s}\Big)^2\Big/v .
\end{equation}
\end{proposition}

With a single cell the corrected moments reduce to the covariate correction of \citet[Appendix~C.2]{LudwigMullainathanRambachan2026LLM}, and the variance~\eqref{eq:variance} to their equation~(15) with a simple random audit; the cells and the stratified design are what the treatment-effect structure adds. Part (i) is what makes the procedure robust to selective errors: it never uses the error \emph{rates} to rescale anything, so it does not need the label's errors to be unrelated to outcomes, which Lemma~\ref{lem:gap} and Table~\ref{tab:delta} show they are not here. Part (iii) is the budget rule: the label sets the full-sample term, and the audit sets a floor that only more audit can lower. For inference I use a two-phase bootstrap that resamples the full sample and, independently, the audited records within their strata; because the audit observes the truth together with the outcome, the bootstrap needs no model of the label's errors. When only the error rates are known, from an external sample without outcomes, the coupled-label bootstrap of \citet{Christensen2026Bootstrap}, which draws true and imputed labels jointly, is the counterpart, under their conditions on the errors; the corrections of \citet{Battaglia2024Inference} are implemented in the \texttt{ValidMLInference} package \citep{ChristensenHansen2026Guide}, and the procedure here in \texttt{llmaudit}. The pilot diagnostic of Step~1 says which applies: if false positives carry the bias and the misses are unselective, a correction from the false-positive rate may be enough. A better label helps through both terms, since it shrinks $S_{\psi,s}$; in the simulations of Section~\ref{sec:size} a label with higher recall cuts the standard error of the corrected ATT by a quarter to a third at a fixed audit size (Section~\ref{sec:size}).

Applying the same difference once to the regression coefficient, as plain prediction-powered inference would, recovers the coefficient one would get with the true treatment. By~\eqref{eq:sloczynski} that coefficient is still a weighted mix of the two groups' effects, with weights set by group sizes. Correcting the cells is what lets the researcher choose the weights.

\subsection{Step 3: choose the estimand}

From the corrected cell moments I compute the ATT, the ATU and the ATE by~\eqref{eq:estimands}, and the coefficient a regression on the true treatment would give, together with its \citeauthor{Sloczynski2022OLS} decomposition $\tilde\rho$, $\tilde w_1$, $\tilde\tau_1$, $\tilde\tau_0$. Reporting the decomposition shows the reader whose effect the regression is about; reporting the cell estimands lets the researcher answer the question actually asked. If the question is what the treatment did where it was adopted, the ATT is the answer; if it is what would happen if it spread, the ATU is. The two estimates of each group's effect, from the linear lines and from the cells, differ only by \citeauthor{Sloczynski2022OLS}'s nonlinearity term, and reporting both is a check on linearity. Confidence intervals come from a two-phase bootstrap that resamples the full sample and, independently, the audited records within their strata.

\section{Application: the remote-work pay gap in LinkedIn postings}
\label{sec:application}

\subsection{Data and labels}

I use 35,618 US job postings from LinkedIn in 2023--24 that report an annualized salary, from the public ``LinkedIn Job Postings (2023--2024)'' collection \citep{Arshkon2024LinkedIn}. The outcome is the log salary. The controls are industry, experience level and work type, coarsened into ten cells of the label's linear propensity score. Each posting carries an employer field marking it as remote (4,787 postings, 13.4 percent), which serves as a benchmark: it is what a researcher with the structured data would use, and it allows a check of the procedure, though the audit shows it is not the truth either. I label each posting from its title and description with three tools: a keyword rule, the DistilBERT classifier of \citet{Hansen2023Remote} used by \citet{Battaglia2024Inference}, and a 7-billion-parameter instruction-tuned language model (Qwen-2.5) asked zero-shot whether the job is fully remote (prompt and settings in Appendix~\ref{app:prompt}). The model never sees the salary column, and its labels do not rise with salary among postings with the same employer field.

Table~\ref{tab:confusion} compares the labels with the employer field. The language model is the most precise and the least complete: it flags 2,450 postings, misses 2,637 of the 4,787 the employer marks remote, and its false discovery rate is 12 percent. The keyword rule and DistilBERT catch more but are wrong about half the time when they flag.

\begin{table}[htbp]\centering\small
\caption{The labels against the employer's remote field}\label{tab:confusion}
\begin{tabular}{lrrrr}\toprule
Label & Flagged & Recall & FDR & Missed \\\midrule
Keyword rule & 5,850 & 0.599 & 0.510 & 1,919 \\
DistilBERT (Hansen et al.) & 5,050 & 0.485 & 0.540 & 2,466 \\
Qwen-2.5-7B, zero-shot & 2,450 & 0.449 & 0.122 & 2,637 \\
Qwen + location rule & 3,784 & 0.720 & 0.089 & 1,338 \\
\bottomrule\end{tabular}
\par\smallskip\parbox{0.9\linewidth}{\footnotesize Notes: 35,618 postings, of which 4,787 (13.4\%) are marked remote by the employer. Recall and false discovery rate (FDR) are measured against the employer field, which is itself audited in Section~\ref{sec:application}. The location rule labels a posting remote when its location is the bare country name.}
\end{table}

\subsection{What the audit found}

I audited 650 postings in four waves, all read blind under the protocol of Section~\ref{sec:procedure}: two readers, outcome masked, cell $\times$ flag strata and a Neyman second wave. The first two waves (200 postings, aimed at the model's apparent false positives and misses) were first read with the salary visible, before I adopted the blind rule; I re-read them blind and use only the blind calls. The comparison is informative in itself, and I return to it below. Six findings shape the correction.

\emph{The benchmark is wrong where the label looks wrong.} Of the language model's apparent false positives, 86 percent are remote according to the posting text: the employer did not tick the box. Against the text, the model's false discovery rate is about 5 percent, not 12. A precision check against the structured field would have condemned the label for the field's errors.

\emph{The bias is in the misses.} Table~\ref{tab:errors} in Section~\ref{sec:bias} showed the pattern against the employer field: false-positive rates under one percent and false discovery rates up to 38 percent where remote is rare; miss rates of 50 to 75 percent in every cell. Table~\ref{tab:whicherror} showed that removing the misses removes the bias. The audit adds what the field cannot. Of the postings the model misses, about 27 percent say they are remote in the text, 51 percent say nothing about the work arrangement, and the rest are hybrid (9 percent) or on-site (14 percent). The text-evident misses are not random: they are longer postings with the remote cue in the middle or at the end, and missed remote postings pay 6.3 log points more than the remote postings the model catches ($t\approx4.3$). These are the selective misses of Lemma~\ref{lem:gap}, and they push the naive estimate down.

\emph{Structured fields carry what the text does not.} A posting whose location is the bare country name, ``United States'', is remote by the employer field 98.6 percent of the time; such postings are 53 percent of all remote postings, and the language model, which reads only the text, flags fewer than half of them. Most of the model's ``silent'' misses have this location. Adding the rule ``country-only location means remote'' to the model's label raises recall from 45 to 72 percent (Table~\ref{tab:confusion}). It is a cheap fix specific to this platform, and the audit supports it: of the 52 audited postings with that location, 22 say remote, 26 say nothing, and only 4 are hybrid or on-site.

\emph{The large unflagged stratum is clean, and that has to be shown.} Postings that the label does not flag and whose text has no remote keyword are 28,500 of the 35,618. None of the 336 read is remote, which bounds the hidden remote rate there below 0.9 percent. Before the second wave the bound was 3.3 percent, enough to move the corrected ATU by 0.03 to 0.08: in that stratum one audited posting stands for about 300.

\emph{Readers disagree, mostly about definitions, and seeing the outcome matters for people.} The two readers agreed on 67 to 73 percent of postings before reconciliation (Cohen's $\kappa$ between 0.53 and 0.55 in the three blind waves). Most disagreements were about what counts as remote (hybrid with home days, remote offered as an option, field jobs with no office) and were settled by one written rule; others were keying slips against explicit text, or calls made from a prior rather than a cue, which the rule codes as undetermined. On a random check of postings the model reader called on-site or uninformative, the person overturned none of its calls that matter for treatment status (0 of 38). Re-reading the first 200 postings blind changed the person's calls in a revealing way: with the salary visible, the postings the person called remote tracked pay about three times as steeply (a slope of 0.14 on log pay, against 0.05 blind), and the postings called remote only when the salary was visible paid more than those called remote only blind (11.61 against 11.21 in log pay). With 200 postings the difference is suggestive rather than conclusive (95 percent interval for the difference in slopes 0.01 to 0.18). The language-model reader, run twice on the same postings with and without the salary, shows no such effect (93 percent agreement; slopes 0.06 and 0.07). Replacing the salary-visible calls by the blind ones lowers the corrected ATT by about 0.013.

\emph{Masking has to be fine-grained.} My first blind sheets deleted every sentence that mentioned pay. That also deleted the treatment when the two shared a sentence (``Location: 100\% Remote. Pay rate: [pay]/hr''), in 6 to 12 percent of postings. Masking only the amounts fixes it; re-reading the affected postings changed four treatment calls.

\subsection{Results}

Table~\ref{tab:linkedin} reports the estimates. The language-model label alone gives a remote pay gap of 0.086 log points; with the location rule, 0.124. After the audit and the cell correction the coefficient is 0.14, the ATT 0.12 to 0.13 and the ATU 0.20 to 0.21, the ranges spanning the two assignments of silent postings. The naive label understates the effect on the treated and the effect on the untreated by 35 to 40 percent, and the corrected estimates land close to the employer-field benchmark (0.150, 0.140, 0.214), which the correction never uses.

\begin{table}[htbp]\centering\small
\caption{The remote-work pay gap: naive, corrected and benchmark}\label{tab:linkedin}
\resizebox{\linewidth}{!}{%
\begin{tabular}{lccccccc}\toprule
 & \multicolumn{3}{c}{Estimand} & \multicolumn{4}{c}{S\l{}oczy\'nski decomposition} \\\cmidrule(lr){2-4}\cmidrule(lr){5-8}
 & $\beta$ & ATT & ATU & $\rho$ & $w_1$ & $\tau_1$ & $\tau_0$ \\\midrule
Naive: Qwen label & 0.086 & 0.081 & 0.124 & 0.069 & 0.917 & 0.076 & 0.192 \\
Naive: Qwen + location rule & 0.124 & 0.114 & 0.174 & 0.106 & 0.849 & 0.104 & 0.237 \\
\addlinespace
Corrected: silent follow field & 0.136 & 0.124 & 0.205 & 0.141 & 0.841 & 0.117 & 0.233 \\
 & [0.075, 0.191] & [0.065, 0.181] & [0.141, 0.273] & & & & \\
Corrected: silent not remote & 0.138 & 0.127 & 0.202 & 0.127 & 0.850 & 0.121 & 0.235 \\
 & [0.082, 0.188] & [0.073, 0.177] & [0.127, 0.273] & & & & \\
Audit only: silent follow field & 0.132 & 0.122 & 0.260 &  &  &  &  \\
Audit only: silent not remote & 0.137 & 0.128 & 0.263 &  &  &  &  \\
\addlinespace
Employer field (benchmark) & 0.150 & 0.140 & 0.214 & 0.134 & 0.808 & 0.127 & 0.248 \\
\bottomrule\end{tabular}}
\par\smallskip\parbox{0.95\linewidth}{\footnotesize Notes: log points. Controls: ten cells of the label's linear propensity score (industry, experience level, work type). $\beta$ is the coefficient on the treatment with cell dummies; $\rho$ the treated share; $w_1$ the weight on the treated; $\tau_1,\tau_0$ the effects on the treated and untreated from the two lines. Corrected rows use 650 audited postings (Section~\ref{sec:application}); 95\% intervals from a two-phase bootstrap (500 draws). ``Silent'' postings are those whose text says nothing about the work arrangement.}
\end{table}

The decomposition columns report \citeauthor{Sloczynski2022OLS}'s $\tau_1$ and $\tau_0$ next to the cell averages ATT and ATU, and the two pairs differ: with the employer field, $\tau_1=0.127$ against an ATT of $0.140$, and $\tau_0=0.248$ against an ATU of $0.214$. The difference is the bias from nonlinearity. $\tau_1$ and $\tau_0$ are read off two straight lines fitted to the cell means against the propensity score, one for each group; the ATT and ATU average the cell means directly. The untreated line is pulled up at high propensity scores by the cells where remote jobs are common and the pay gap is small, which is where the treated sit, so the line understates their counterfactual and $\tau_1$ falls below the ATT; the mirror image raises $\tau_0$ above the ATU. \citet{Sloczynski2022OLS} finds the same pattern in his applications, with linearity holding for the treated but not the untreated. Reporting both pairs shows the reader how much the linear reading of the regression can be trusted here; the cell averages are the estimands I recommend. The decomposition columns also show what the correction does to whose effect the regression reports. The naive label halves the treated share and pushes the weight on the treated from 0.81 to 0.92, so the naive coefficient is almost entirely an attenuated ATT. After correction the treated share is 13 to 14 percent and $w_1=0.84$ to $0.85$: the regression is still mostly about remote postings, as \citet{Sloczynski2022OLS} predicts when the treated are the minority. But the two effects differ. The pay gap for postings that are remote (0.12--0.13) is about 40 percent smaller than the gap that would apply to postings that are not (0.20--0.21), so a reader who takes the regression coefficient as the effect of spreading remote work would understate it. The audit-only estimates, which use the 650 audited postings and no labels, agree on the ATT but put the ATU higher; with 650 postings, the audit alone has only a handful of remote postings in the low-propensity cells that drive the ATU, which is where the label's full-sample information helps most.

\paragraph{Comparison with \citet{Battaglia2024Inference}.} Their application asks the same question with different data and a different label. In the current version of their paper (arXiv v5, April 2025) they regress log posted wages on a remote indicator for 16,315 Lightcast postings in accommodation and food services (NAICS 72) in San Diego in 2022--23, with remote status imputed by the DistilBERT classifier of \citet{Hansen2023Remote}. Their bias correction raises the coefficient by 62 percent without controls (from 0.648 to 1.052) and by 76 percent with two-digit occupation effects (from 0.364 to 0.641). My correction moves the language-model estimate by a similar proportion, about 60 percent. The two corrections rest on different measurements, though. Theirs is driven by the label's false-positive rate, which they estimate by drawing 1,000 postings at random and reading the 26 the classifier flags: nine are not remote, a false-positive rate of 0.009. This is the precision check of Section~\ref{sec:measure}, and it shows the base-rate arithmetic of Section~\ref{sec:label} at work: a false-positive rate under one percent leaves about a third of the flagged column (9 of 26) contaminated, close to the 38 percent in cell 1 of the LinkedIn data (Table~\ref{tab:twocells}). Their audit does not read the 974 unflagged postings, so it cannot measure misses. \citet[p.~94]{ChristensenHansen2026Guide} state the logic plainly: because only 26 of the 1,000 sampled postings are predicted remote, ``only those require inspection.'' Under their assumptions that is right; the question this paper asks is whether those assumptions hold, and only the unflagged postings can answer it. \citet{Christensen2026Bootstrap}, revisiting the same San Diego regressions with their bootstrap, set the false-negative rate equal to the false-positive rate and double it as a robustness check; their implementation of the analytic correction gives 0.90 rather than 1.05 without controls, and I quote the figures of \citet{Battaglia2024Inference}. The LinkedIn data suggest that for a language-model label the misses can be many times the false positives (Table~\ref{tab:confusion}). In the LinkedIn data the misses are the bias: removing them removes all of it, while removing the false positives removes none (Table~\ref{tab:whicherror}). The same Hansen classifier, run on the LinkedIn postings, misses half of the employer-remote postings and has a false discovery rate of 54 percent against the employer field (Table~\ref{tab:confusion}), in part because its human labels count a job as remote when the posting ``explicitly mentions the possibility to work from home at least one day per week'' \citep[p.~94]{ChristensenHansen2026Guide}, which includes hybrid. The lesson is that which error matters depends on the label and on the definition of the treatment, and that a researcher cannot know which case they are in without reading the unflagged column: a check of flagged records is enough when the label's errors are false positives, and blind to the bias when they are misses.

\begin{figure}[htbp]\centering
\includegraphics[width=0.85\linewidth]{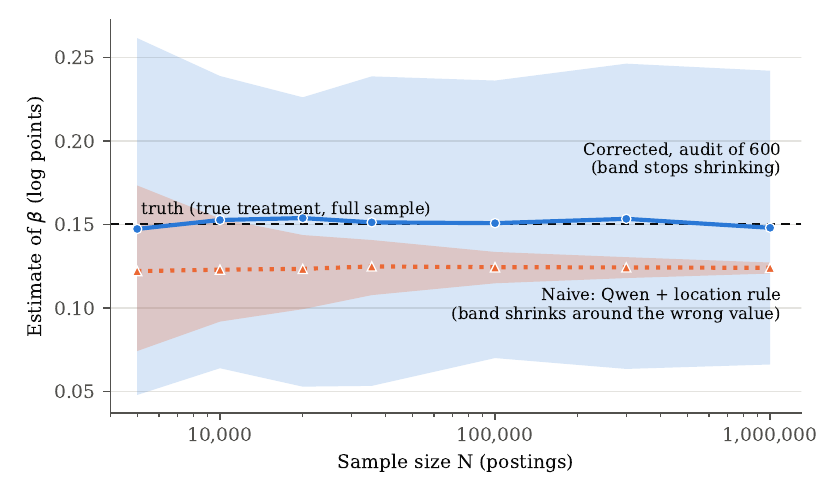}
\caption{Naive and corrected estimates as the sample grows, with a fixed audit}\label{fig:floor}
\par\smallskip\parbox{0.9\linewidth}{\footnotesize Notes: lines are means and bands the 2.5th to 97.5th percentiles across 200 samples of size $N$ drawn with replacement from the 35,618 postings. Truth: the employer's remote field. Naive: the label with the location rule. Corrected: the same label with an audit of 600 postings per sample, stratified by cell $\times$ flag.}
\end{figure}

Figure~\ref{fig:floor} shows what the audit buys and what it does not. As the sample grows, the naive estimate's sampling band shrinks around 0.124, below the truth; the corrected estimate stays centred on the truth, but with a fixed audit of 600 postings its band stops shrinking once the audit term of~\eqref{eq:variance} dominates. A larger language-model sample does not narrow it; a larger audit or a better label does.

\section{How large an audit}
\label{sec:size}

To see how the corrected estimates behave as the audit grows, I treat the employer field as the truth, which is known for every posting, and draw repeated pseudo-audits of size $m$ from the cell $\times$ flag strata. Table~\ref{tab:size} reports the standard deviation of the ATT and ATU across 150 audits for each design, for the language-model label alone and with the location rule, under proportional and Neyman allocation, and compares the corrected estimator with using the audit alone.

\begin{table}[htbp]\centering\small
\caption{How large an audit: standard deviation of the estimates across simulated audits}\label{tab:size}
\begin{tabular}{llrcccc}\toprule
 & & & \multicolumn{2}{c}{ATT} & \multicolumn{2}{c}{ATU} \\\cmidrule(lr){4-5}\cmidrule(lr){6-7}
Label & Allocation & Audit $m$ & Audit only & Corrected & Audit only & Corrected \\\midrule
Qwen + location & neyman & 300 & 0.091 & 0.069 & 0.088 & 0.069 \\
Qwen + location & neyman & 600 & 0.067 & 0.044 & 0.069 & 0.060 \\
Qwen + location & neyman & 1,200 & 0.043 & 0.031 & 0.055 & 0.048 \\
\addlinespace
Qwen + location & proportional & 300 & 0.081 & 0.071 & 0.081 & 0.057 \\
Qwen + location & proportional & 600 & 0.066 & 0.049 & 0.069 & 0.059 \\
Qwen + location & proportional & 1,200 & 0.043 & 0.033 & 0.051 & 0.046 \\
\addlinespace
Qwen & neyman & 300 & 0.103 & 0.082 & 0.104 & 0.091 \\
Qwen & neyman & 600 & 0.062 & 0.052 & 0.087 & 0.072 \\
Qwen & neyman & 1,200 & 0.044 & 0.040 & 0.062 & 0.056 \\
\addlinespace
Qwen & proportional & 300 & 0.093 & 0.098 & 0.109 & 0.104 \\
Qwen & proportional & 600 & 0.071 & 0.068 & 0.096 & 0.093 \\
Qwen & proportional & 1,200 & 0.043 & 0.045 & 0.060 & 0.061 \\
\bottomrule\end{tabular}
\par\smallskip\parbox{0.95\linewidth}{\footnotesize Notes: truth = the employer's remote field, known for all 35,618 postings; 150 simulated audits per row. Proportional: a third of the audit to flagged strata, the rest in proportion to stratum size. Neyman: 30\% proportional pilot, the rest by Neyman allocation with rule-of-three floors. Bias of the corrected ATT is below 0.015 in every row; the corrected ATU is biased down by up to 0.04 with the Qwen label at small audits, where cells with few treated units make the ratio unstable.}
\end{table}

Three lessons follow. First, the correction removes the bias at every audit size considered; the exception is the ATU with the weaker label and a small audit, where cells with few treated postings make the ratio unstable. Second, how much the label's full-sample information buys depends on the label. With the language-model label alone and proportional allocation, the corrected ATT is no more precise than the audit alone; the label helps only through the design, when Neyman allocation uses it to decide where to read. With the location rule, which raises recall from 45 to 72 percent, the corrected ATT has a standard deviation about a quarter smaller than the audit alone (12 to 34 percent across rows), worth roughly an audit 1.8 times as large. Third, Neyman allocation helps most when the label is weak: it cuts the variance of the corrected ATT by 21 to 42 percent with the language model alone, against 6 to 19 percent with the better label, because a weak label leaves more of the error in strata a proportional design under-reads.

The budget rule~\eqref{eq:budget} turns these numbers into a planning tool. Across rows the standard deviation of the corrected ATT is close to $c/\sqrt m$, with $c\approx1.1$ to $1.2$ for the label with the location rule and $c\approx1.4$ to $1.6$ for the language model alone. A researcher who wants a standard error of 0.03 log points on the ATT therefore needs an audit of about $(1.15/0.03)^2\approx1{,}500$ postings with the better label and 2,100 to 3,000 with the weaker one; for 0.05, about 550 and 750 to 1,100. Because the audit term does not shrink with $N$, these numbers do not depend on how many postings the language model labels. In practice $c$ comes from the pilot through $S_{\psi,s}$, and the second wave is sized to reach the target.

\section{Conclusion}
\label{sec:conclusion}

A language-model label makes a regression cheap to run on a very large sample, and that is exactly when its errors matter most: the bias stays while the standard error vanishes. Under heterogeneous effects the damage is twofold, because the label changes both the group effects and the weights OLS places on them. The remedy is not a better test of the label's accuracy but a measurement of what the label gets wrong where it matters: audit the label, correct the cells, choose the estimand.

In practice, a researcher with an LLM-labeled treatment would proceed in five steps: write down the treatment rule, including the ambiguous cases; look for structured fields that already carry the treatment; bin the label's propensity score into cells; read a blind pilot of a few hundred records drawn from every cell and from both the flagged and the unflagged columns, with a language model and a person as the two readers; and follow it with a second wave allocated by Neyman's rule and sized by the budget rule. The cell-by-cell correction then gives the effect on the treated, the effect on the untreated and the regression coefficient with its weights, and the researcher reports the one that answers the question. The \texttt{llmaudit} package implements these steps.

The procedure rests on ignorability within cells and on cells that are coarser than the full set of controls; it estimates the effect of the treatment as the text reveals it, which need not equal a structured field, and it bounds rather than resolves records whose text is silent. The audit's readers are themselves a source of error, which is why the protocol uses two readers, a written rule and a reader check, and why the readers must never see the outcome.

\bibliography{refs}

\newpage
\appendix
\section{Proofs}
\label{app:proofs}

\paragraph{Lemma~\ref{lem:gap}: derivation.} Fix a cell $w$ and drop the argument $w$ throughout. Recall the notation: $D$ is the true treatment and $\Dh$ the label; $a=\Pr(\Dh=1\mid D=0)$ is the false-positive rate and $m=\Pr(\Dh=0\mid D=1)$ the miss rate; $\FDR=\Pr(D=0\mid\Dh=1)$ and $\FOR=\Pr(D=1\mid\Dh=0)$ are the contaminations of the flagged and unflagged columns; $\mu_d=\E[Y\mid D=d]$ and $\mu_{d\ell}=\E[Y\mid D=d,\Dh=\ell]$; $\Delta_1=\mu_{11}-\mu_{10}$ and $\Delta_0=\mu_{01}-\mu_{00}$.

\emph{Step 1: each column mean is a mixture over true status.} By the law of total expectation, conditioning on $D$ within each column,
\begin{align*}
\E[Y\mid\Dh=1]&=\Pr(D=1\mid\Dh=1)\,\mu_{11}+\Pr(D=0\mid\Dh=1)\,\mu_{01}=(1-\FDR)\,\mu_{11}+\FDR\,\mu_{01},\\
\E[Y\mid\Dh=0]&=\Pr(D=1\mid\Dh=0)\,\mu_{10}+\Pr(D=0\mid\Dh=0)\,\mu_{00}=\FOR\,\mu_{10}+(1-\FOR)\,\mu_{00}.
\end{align*}

\emph{Step 2: each group mean is a mixture over the label.} By the same law, conditioning on $\Dh$ within each true group,
\[
\mu_1=(1-m)\,\mu_{11}+m\,\mu_{10},\qquad \mu_0=a\,\mu_{01}+(1-a)\,\mu_{00}.
\]

\emph{Step 3: express the four cell-by-label means through the group means and the selectivity gaps.} Solving the two equations of Step~2 together with the definitions of $\Delta_1$ and $\Delta_0$,
\[
\mu_{11}=\mu_1+m\,\Delta_1,\quad \mu_{10}=\mu_1-(1-m)\,\Delta_1,\quad
\mu_{01}=\mu_0+(1-a)\,\Delta_0,\quad \mu_{00}=\mu_0-a\,\Delta_0 .
\]
(Check the first: $(1-m)\mu_{11}+m\mu_{10}=(1-m)(\mu_1+m\Delta_1)+m(\mu_1-(1-m)\Delta_1)=\mu_1$.) In words: the treated units the label catches earn the group mean plus $m\Delta_1$, and the ones it misses earn the group mean minus $(1-m)\Delta_1$; if $\Delta_1=0$ both earn the group mean.

\emph{Step 4: substitute into Step 1 and subtract.}
\begin{align*}
\hat\tau&=\E[Y\mid\Dh=1]-\E[Y\mid\Dh=0]\\
&=(1-\FDR)(\mu_1+m\Delta_1)+\FDR\,\{\mu_0+(1-a)\Delta_0\}\\
&\qquad-\FOR\,\{\mu_1-(1-m)\Delta_1\}-(1-\FOR)(\mu_0-a\Delta_0)\\
&=(1-\FDR-\FOR)(\mu_1-\mu_0)\\
&\qquad+\big[(1-\FDR)\,m+\FOR\,(1-m)\big]\Delta_1+\big[\FDR\,(1-a)+(1-\FOR)\,a\big]\Delta_0 ,
\end{align*}
which is~\eqref{eq:gap} with the weights~\eqref{eq:kappa}; the weights are non-negative because every factor is a probability. Nothing so far uses Assumption~\ref{as:ign2}: the identity holds for any binary variable $D$ and label $\Dh$. Under Assumption~\ref{as:ign2}, $\mu_d=\E[Y(d)]$ within the cell, so $\mu_1-\mu_0=\tau(w)$, the causal effect in the cell.

\emph{Step 5: the classical case.} The label's errors are non-differential when $\Dh\perp Y\mid(D,W)$: within a cell and a true status, the label carries no information about the outcome. Then $\mu_{11}=\mu_{10}=\mu_1$ and $\mu_{01}=\mu_{00}=\mu_0$, so $\Delta_1=\Delta_0=0$ and $\hat\tau=(1-\FDR-\FOR)\tau$. With a single cell (no controls) and a common effect this is \citeauthor{Aigner1973Binary}'s (\citeyear{Aigner1973Binary}) attenuation of the coefficient on a misclassified binary regressor, expressed through the contaminations of the two columns rather than the classifier's error rates; the two forms are linked by~\eqref{eq:fdrfor}. \qed

\paragraph{Proposition~\ref{prop:channels}.} With cell dummies as the controls, the linear propensity score of any binary regressor is its cell mean, so Theorem~1 of \citet{Sloczynski2022OLS} applies both to $D$ (giving~\eqref{eq:sloczynski}) and to $\Dh$ (giving $\hat\beta=\hat w_1\hat\tau_1+\hat w_0\hat\tau_0$ with all quantities computed on the labeled columns). Subtract and use $\hat w_0-w_0=-(\hat w_1-w_1)$:
$\hat\beta-\beta=\hat w_1(\hat\tau_1-\tau_1)+\hat w_0(\hat\tau_0-\tau_0)+(\hat w_1-w_1)\tau_1+(\hat w_0-w_0)\tau_0$, and the last two terms equal $(\hat w_1-w_1)(\tau_1-\tau_0)$. For the rare-treatment form, let $\hat\rho\to0$ with $\FOR(w)\to0$: then $\hat w_1\to1$, $\hat\tau_1$ averages $\hat\tau(w)$ over the flagged column, and by Lemma~\ref{lem:gap} with $\delta=0$ it equals $(1-\FDRbar)\ATT$ up to terms that vanish with $\FOR$. \qed

\paragraph{Proposition~\ref{prop:regime}: derivation.} Use the notation of Lemma~\ref{lem:gap} in one cell: $\mu_1=(1-m)\mu_{11}+m\mu_{10}$, $\mu_0=a\mu_{01}+(1-a)\mu_{00}$, so $\mu_{11}=\mu_1+m\Delta_1$, $\mu_{10}=\mu_1-(1-m)\Delta_1$, $\mu_{01}=\mu_0+(1-a)\Delta_0$ and $\mu_{00}=\mu_0-a\Delta_0$.

\emph{Misses only.} Set the false positives to their true value 0, keeping the truth fixed. The flagged column holds the caught treated units, with mean $\mu_{11}$; the unflagged column holds the missed treated units and all untreated units, a share $\FOR_M=pm/(pm+1-p)$ of it missed, with mean $\FOR_M\mu_{10}+(1-\FOR_M)\mu_0$. Then
\begin{align*}
\hat\tau_M&=\mu_1+m\Delta_1-\FOR_M\{\mu_1-(1-m)\Delta_1\}-(1-\FOR_M)\mu_0\\
&=(1-\FOR_M)\tau+\{m+\FOR_M(1-m)\}\Delta_1,
\end{align*}
so $b_M=\hat\tau_M-\tau=-\FOR_M\tau+\{m+\FOR_M(1-m)\}\Delta_1$, which divided by $\tau$ is~\eqref{eq:bM}.

\emph{False positives only.} Set the misses to their true value 1. The flagged column holds all treated units and the false positives, a share $\FDR_F=(1-p)a/\{p+(1-p)a\}$ of it; the unflagged column holds the true negatives, with mean $\mu_{00}$. Then
\begin{align*}
\hat\tau_F&=(1-\FDR_F)\mu_1+\FDR_F\{\mu_0+(1-a)\Delta_0\}-\mu_0+a\Delta_0\\
&=(1-\FDR_F)\tau+\{a+\FDR_F(1-a)\}\Delta_0,
\end{align*}
which divided by $\tau$ is~\eqref{eq:bF}. Substituting $(1-p)a=pm/R$ gives $\FDR_F=m/(m+R)$.

\emph{Interaction and the flagged share.} $b_{MF}$ is defined as $\hat\tau-\tau-b_M-b_F$, with $\hat\tau$ from Lemma~\ref{lem:gap}. The flagged share is $\hat p=p(1-m)+(1-p)a$, so $\hat p-p=(1-p)a-pm=pm(1/R-1)$, negative exactly when $R>1$. \qed

\paragraph{The grid behind Figure~\ref{fig:regime}.} The boundaries solve $|b_M|=|b_F|$ on a grid of 120 base rates $p$ from 0.5 to 50 percent and 120 values of $R$ from 0.1 to 50, with $a=pm/\{(1-p)R\}$. The figure uses $m=0.5$. The boundaries move little with the miss rate: at $p=3$ percent and $\delta_0=0$, the value of $R$ above which misses carry more of the bias is 33 for all $m$ when $\delta_1=0$; 3.3, 3.1 and 2.8 for $m=0.25$, 0.5 and 0.75 when $\delta_1=-0.25$; and 1.7, 1.4 and 1.2 when $\delta_1=-0.5$. Applied to the ten LinkedIn cells with their own $p$, $a$, $m$, $\tau$, $\Delta_1$ and $\Delta_0$, Lemma~\ref{lem:gap} reproduces each cell's labeled gap to machine precision. The code is \texttt{sims/regime\_map.py} and \texttt{sims/regime\_plot.py}.

\paragraph{Coverage~\eqref{eq:coverage}.} $\sqrt N(\hat\beta_N-\hat\beta)/\sigma\to\mathcal N(0,1)$, and the interval $\hat\beta_N\pm1.96\sigma/\sqrt N$ contains $\beta$ iff $|\sqrt N(\hat\beta_N-\hat\beta)/\sigma+r_N|\le1.96$. \qed

\paragraph{Proposition~\ref{prop:audit}.} (i) Conditional on the full sample, the first terms of~\eqref{eq:corrected} are fixed and the second are Horvitz--Thompson estimators of $N_w^{-1}\sum_{i\in w}Y_i(D_i-\Dh_i)$ and $N_w^{-1}\sum_{i\in w}(D_i-\Dh_i)$, unbiased whenever $\pi_i>0$. Their sums are therefore unbiased for the full-sample true moments $N_w^{-1}\sum_{i\in w}Y_iD_i$ and $N_w^{-1}\sum_{i\in w}D_i$, with no condition on how $D-\Dh$ relates to $Y$. (ii) Each estimand in~\eqref{eq:estimands}, and $\beta$, is a continuous function of the cell moments on the set where $0<p(w)<1$; consistency follows from the consistency of stratified Horvitz--Thompson means as $n_s\to\infty$ and of full-sample means as $N\to\infty$. (iii) Linearize $\tilde\theta$ around the true moments. The full-sample and audit terms contribute separately (the audit is drawn conditionally on the full sample), giving $\sigma_1^2/N$ plus the stratified-sampling variance of the influence $\psi_i=\nabla\theta\cdot(Y_i,1)'(D_i-\Dh_i)$, which is the second term of~\eqref{eq:variance}. Minimizing $\sum_sW_s^2S_{\psi,s}^2/n_s$ subject to $\sum_sn_s=n$ gives $n_s\propto W_sS_{\psi,s}$ and minimum $(\sum_sW_sS_{\psi,s})^2/n$ \citep{Neyman1934Two}; setting it equal to $v$ gives~\eqref{eq:budget}. \qed

\section{The language-model label and the reader protocol}
\label{app:prompt}

\paragraph{The label.} Each posting is labeled by Qwen-2.5-7B-Instruct, an open-weight instruction-tuned model with 7 billion parameters, run locally in 8-bit precision (the \nolinkurl{mlx-community/Qwen2.5-7B-Instruct-8bit} weights). The model reads the posting's title and description, never the salary field or the employer's remote field, and descriptions longer than 2,000 tokens are cut at 2,000 (about 1 percent of postings). I do not let the model generate text. I read the probability it assigns to ``Yes'' and to ``No'' as the next token and flag the posting when the probability of ``Yes'' exceeds one half. The system message is:
\begin{quote}\small\ttfamily
You classify job postings. Answer with exactly one word: Yes or No.
\end{quote}
and the user message is:
\begin{quote}\small\ttfamily\raggedright
Read the job posting below and decide whether the job is a REMOTE position: the employee can do the job entirely from home or another location of their choosing, without regularly reporting to an employer site.\\
Answer Yes only if the posting indicates that the job is fully remote.\\
Answer No if the job is on-site, hybrid (some days in the office), tied to a specific work site, or if the posting does not indicate that the job is remote.\\[4pt]
Job posting:\\
"""\\
\{title and description\}\\
"""\\[4pt]
Is this a remote position? Answer Yes or No.
\end{quote}

\paragraph{The audit readers.} Every audited posting is read twice, independently, by me and by a language model (Claude, Anthropic), from a blind sheet: pay amounts are replaced by ``[pay]'', the salary field, the label and the sampling stratum are not shown, and postings are shuffled. Both readers apply the same written rule and record a verdict, a confidence level and the deciding phrase verbatim. The rule, as finally agreed, is:
\begin{itemize}\small
\item \emph{Remote}: the employee does not regularly report to an employer site. This includes fully remote and work-from-home jobs, home-based jobs with occasional travel or training, ``locally remote'' jobs that require living in a region, jobs that offer remote work as an option, and jobs whose location line reads ``Remote''.
\item \emph{Hybrid}: a regular mix of office days and home days.
\item \emph{On-site}: the text states or clearly implies work at a specific site. Field jobs with no employer site (field sales in a territory, home-health visits) are coded on-site and flagged.
\item \emph{Not about the job}: the word ``remote'' or ``virtual'' appears but not about where the employee works.
\item \emph{No information}: the text says nothing usable about the arrangement. A city or state alone, and company-wide boilerplate (``we are a remote-first company'', benefits lists), are no information. A reader's prior about the job is not evidence: an on-site call needs a cue in the text.
\end{itemize}
Disagreements are resolved by applying the rule, and both original calls are kept. The model reader's instructions are reproduced in the replication package together with the \texttt{llmaudit} software.

\end{document}